\documentclass[apj]{emulateapj}

\newcommand{\kms}{km~s$^{-1}$} 
\newcommand{\mum}{$\mu$m} 
 
\newcommand{\methanol}{CH$_3$OH}

\newcommand{\lsun}{$L_\odot$}
\newcommand{\msun}{$M_\odot$}

\newcommand{\ngc}{NGC6334}
\newcommand{\ngci}{{\ngc}I}

\newcommand{\mjb}{mJy~beam$^{-1}$}

\newcommand{\HII}{H\,{\sc ii}}

\newcommand{\ccc}{cm$^{-3}$}
\newcommand{\Elower}{$E_{Lower}\widehat{=}$}
\newcommand{\tdust}{$T_{dust}$}

\slugcomment{submitted to ApJ on November 13, 2017; in revised form on January 4, 2018; accepted on January 6, 2018}

\shorttitle{Extraordinary Outburst in NGC6334I-MM1: \methanol\/ masers}
\shortauthors{Hunter et al.}

\begin{document}

\title{ The extraordinary outburst in the massive protostellar system NGC6334I-MM1:\\ emergence of strong 6.7 GHz methanol masers }


\author{
  T. R. Hunter\altaffilmark{1},
  C. L. Brogan\altaffilmark{1}, 
  G. C. MacLeod\altaffilmark{2}, 
  C. J. Cyganowski\altaffilmark{3}, 
  J. O. Chibueze\altaffilmark{4,5,6}, 
  R. Friesen\altaffilmark{1,7}, 
  T. Hirota\altaffilmark{8}, 
  D. P. Smits\altaffilmark{9},  
  C. J. Chandler\altaffilmark{10}, 
  R. Indebetouw\altaffilmark{1,11}
  }
 
\email{thunter@nrao.edu}

\altaffiltext{1}{NRAO, 520 Edgemont Rd, Charlottesville, VA 22903, USA}  
\altaffiltext{2}{Hartebeesthoek Radio Astronomy Observatory, PO Box 443, Krugersdorp 1740, South Africa}
\altaffiltext{3}{SUPA, School of Physics and Astronomy, University of St. Andrews, North Haugh, St. Andrews KY16 9SS, UK} 
\altaffiltext{4}{Department of Physics and Astronomy, University of Nigeria, Carver Building, 1 University Road, Nsukka, Nigeria}
\altaffiltext{5}{SKA South Africa, 3rd Floor, The Park, Park Road, Pinelands, Cape Town, 7405, South Africa}
\altaffiltext{6}{Centre for Space Research, Physics Department, North-West University, Potchefstroom, 2520, South Africa}
\altaffiltext{7}{Dunlap Institute for Astronomy \& Astrophysics, University of Toronto, 50 St. George St., Toronto, Ontario, Canada, M5S 3H4}  
\altaffiltext{8}{National Astronomical Observatory of Japan, Osawa 2-21-1, Mitaka, Tokyo 181-8588, Japan}
\altaffiltext{9}{Dept of Mathematical Sciences, UNISA, PO Box 392, UNISA, 0003, South Africa}
\altaffiltext{10}{NRAO, PO Box O, Socorro, NM 87801, USA}
\altaffiltext{11}{Department of Astronomy, University of Virginia, P.O. Box 3818, Charlottesville, VA 22903, USA} 
\begin{abstract}
We report the first sub-arcsecond VLA imaging of 6~GHz continuum, methanol maser, and excited-state hydroxyl maser emission toward the massive protostellar cluster NGC6334I following the recent 2015 outburst in (sub)millimeter continuum toward MM1, the strongest (sub)millimeter source in the protocluster.  In addition to detections toward the previously known 6.7~GHz Class~II methanol maser sites in the hot core MM2 and the UCHII region MM3 (NGC6334F), we find new maser features toward several components of MM1, along with weaker features $\sim1''$ north, west, and southwest of MM1, and toward the non-thermal radio continuum source CM2.  None of these areas have heretofore exhibited Class~II methanol maser emission in three decades of observations.  The strongest MM1 masers trace a dust cavity, while no masers are seen toward the strongest dust sources MM1A, 1B and 1D.  The locations of the masers are consistent with a combination of increased radiative pumping due to elevated dust grain temperature following the outburst, the presence of infrared photon propagation cavities, and the presence of high methanol column densities as indicated by ALMA images of thermal transitions.  The non-thermal {radio emission} source CM2 ($2''$ north of MM1) also exhibits new maser emission from the excited 6.035 and 6.030~GHz OH lines.  Using the Zeeman effect, we measure a line-of-sight magnetic field of +0.5 to +3.7~mG toward CM2. In agreement with previous studies, we also detect numerous methanol and excited OH maser spots toward the UCHII region MM3, with predominantly negative line-of-sight magnetic field strengths of $-2$ to $-5$~mG and an intriguing south-north field reversal.
\end{abstract}


\section{Introduction}

\begin{deluxetable*}{lcc}   
\tablewidth{0pc}
\tablecaption{VLA observing parameters\label{obs}}  
\tablehead{\colhead{Parameter} & \colhead{C band} & \colhead{K band} }
\startdata
Project code          & 16B-402   & 10C-186 \\
Observation date(s)   & 2016 Oct 29, 2016 Nov 19 & 2011 May 23\\
Mean epoch            & 2016.9    & 2011.4 \\
Configuration         & A         & BnA \\
Time on source (min)  & 58, 58    & 92 \\
Number of antennas    &  26, 27   & 27 \\
FWHM primary beam ($\arcmin$) & 7 & 2 \\
Baseband center frequencies (GHz)  & 5, 7 & 24.059, 25.406\\
Polarization products & dual circular & dual circular\\ 
Gain calibrator          & J1717-3342 & J1717-3342\\
Bandpass calibrator      & J1924-2914 & J1924-2914\\
Flux calibrator          & 3C286    & 3C286 \\

Spectral windows         & 32       &  16 \\
Digitizer resolution     & 3-bit    & 8-bit\\
Channel spacing (narrow, wide kHz)    & 1.953, 1000      & 31.25\\
Total bandwidth (GHz)    & 3.9      & 0.128 \\
Proj. baseline range (k$\lambda$) & 4.6 - 980 & 12 - 1030 \\
Robust clean parameter & $-1.0$ & +1.0 \\  
Cont. Resolution ($\arcsec\times\arcsec$ (P.A.$\arcdeg$)) & $0.63\times 0.14$ ($-2.9$) & ... \\
Cont. RMS noise (\mjb\/)\tablenotemark{\label{foo}a} & 0.022 & ...\\
Line Resolution ($\arcsec\times\arcsec$ (P.A.$\arcdeg$)) & $0.76\times 0.21$($-3.2$)\tablenotemark{b} [6.7 GHz] & $0.35\times 0.26$($+3.3$) [24.51 GHz] \\
Line channel width (\kms\/)     & 0.15    & 2.0 \\
RMS noise per channel (\mjb\/) & 2.2\tablenotemark{c} [6.7 GHz] & 0.5
\enddata
\tablenotetext{a}{The continuum rms noise varies significantly with position in the image due to dynamic range limitations; the number provided here is a representative value measured near the sources.}
\tablenotetext{b}{The 6~cm excited OH cubes have slightly larger beams due to their lower frequency: $0\farcs79\times 0\farcs25$($-3.2\arcdeg$).}
\tablenotetext{c}{The value reported is for Stokes I in a line-free channel, however, the spectral line sensitivity is significantly worse in channels with strong signal due to dynamic range effects. The signal-free rms noise for the RCP and LCP excited OH cubes is a factor of 1.414 times larger.}
\end{deluxetable*}

Episodic accretion in protostars is increasingly recognized as being an essential phenomenon in star formation \citep{Kenyon90,Evans09}.  The total luminosity of a protostar scales with the instantaneous accretion rate, so variations in that rate will lead to observable brightness changes \citep[e.g.,][]{Offner11}.  The classical manifestations of these changes { in low-mass protostars} are the FU~Orionis stars \citep[FUors,][]{Herbig77}, which exhibit optical flares of 5 or more magnitudes followed by a slow decay, and the EX~Lupi stars \citep[EXors,][]{Herbig89}, which exhibit smaller, 2-3 magnitude flares but in many cases have been seen to repeat.
In recent years, near-infrared surveys such as the Vista Variables in the Via Lactea \citep[VVV,][]{Minniti10} and the United Kingdom Infrared Deep Sky Survey \citep[UKIDSS,][]{Lucas08} have identified hundreds of high-amplitude variables, most of which have been shown to be protostars in earlier evolutionary states \citep[Class~I and II,][]{Contreras17,Lucas17}.  Outbursts in even younger and more deeply-embedded low-mass protostars have recently been detected via large increases in their mid-infrared or submillimeter emission, including a Class~0 source \citep[HOPS-383,][]{Safron15} from the Herschel Orion Protostar Survey \citep[HOPS, e.g.,][and references therein]{Stutz13} and a Class~I source \citep[EC53,][]{Yoo17} from the James Clerk Maxwell Telescope (JCMT) Transient Survey \citep{Mairs17}. 

{ In more massive protostars, indirect evidence for episodic accretion has been inferred from outflow features seen toward high mass young stellar objects (HMYSOs).  Examples include interferometric millimeter CO images of massive bipolar outflows that show high velocity bullets \citep[e.g.][]{Qiu09}, and near-infrared CO spectra of HMYSOs that show multiple blue-shifted absorption features analogous to the absorption features seen in FUOrs \citep[e.g.,][]{Ellerbroek11,Thi10,Mitchell91}.}
Recently, it has become { more directly} clear that massive protostars also exhibit outbursts as evidenced by the 4000~\lsun\/ erratic variable V723~Carinae \citep{Tapia15}, the infrared flare from the 20~\msun\/ protostar powering S255IR-NIRS3 \citep{Caratti16}, and the ongoing (sub)millimeter flare of the deeply-embedded source \ngci-MM1 \citep{Hunter17}. The large increases in bolometric luminosity observed in these events provide evidence for accretion outbursts similar to those predicted by hydrodynamic simulations of massive star formation \citep{Meyer17}. Identifying additional phenomena associated with these events will help to explore the mechanism of the outbursts, particularly if spectral line tracers can be identified, as they can potentially trace gas motions at high angular resolution.

Among the strongest molecular lines emitted from regions of massive star formation are a number of maser transitions \citep{Reid81,Elitzur92}.  Ever since its discovery, maser emission { from protostars} has exhibited eruptive phenomena, such as the three past outbursts observed in the 22~GHz water masers in Orion~KL \citep{Abraham81,Omodaka99,Hirota14}, multiple events in W49N \citep{Honma04,Liljestrom00}, and most recently the outburst in G25.65+1.05 \citep{Volvach17a,Volvach17b,Lekht17}.  The repeating nature of the Orion maser outbursts \citep{Tolmachev11}, as well as the periodic features seen toward many { HMYSOs}
in the 6.7~GHz methanol maser line \citep[e.g.,][]{Goedhart04}, strongly suggest that variations in the underlying protostar could be responsible for the changes in maser emission.  { Such a link was suggested by \citet{Fujisawa12} to explain the methanol maser flare in the HMYSO G33.64-0.21.}  While \citet{Kumar16} noted a similarity between the infrared light curves of several low-mass protostars in the VVV survey and the periodic methanol maser light curves of { HMYSOs}, these phenomena have never been observed in the same object.  However, the recent methanol maser flare in S255IR \citep{Fujisawa15} associated with the infrared outburst in S255IR-NIRS3 has provided the first direct link between protostellar accretion outbursts and maser flares \citep{Moscadelli17}.  


Beginning in early 2015, a strong flaring of many species of masers toward the millimeter protocluster \ngci\/ \citep{Brogan16,Hunter06} was discovered via a regular program of single-dish maser monitoring at the Hartebeesthoek Radio Observatory \citep{MacLeod17}.  Of the four major millimeter sources in this deeply-embedded protocluster, only the UCHII region MM3 \citep[also known as NGC6334F,][]{Rodriguez82} is detected { in the near- and mid-}infrared { \citep[e.g.,][]{Willis13,Walsh99,Tapia96}, despite observations at wavelengths} as long as 18~\mum\/ \citep{DeBuizer02,Persi98}. 
{ While this protocluster is associated with a high-velocity, 0.5~pc-scale NE/SW bipolar outflow  \citep{Zhang14,Qiu11,Beuther08,Leurini06,Ridge01,McCutcheon00}, it remains uncertain which protostar is the driving source; a component
within MM1 seems the most likely, especially considering the presence of water masers there \citep{Brogan16}. }
The fact that the velocity range of the flaring masers encompassed the LSR velocity of the hot molecular core thermal gas \citep[$-$7.2~\kms,][]{McGuire17,Zernickel12,Beuther07} provided strong evidence for an association { between the flaring masers and} the (sub)millimeter continuum outburst in the MM1 hot core \citep{Hunter17}. However, the angular resolution of the maser observations was insufficient to be certain.  

In this paper, we present follow-up observations of \ngci\/ with the National Science Foundation's Karl G. Jansky Very Large Array (VLA) in 5~cm continuum, the 6.7~GHz \methanol\/ transition, and multiple transitions of  excited-state OH.  The relatively nearby distance to the source \citep[1.3$\pm$0.1~kpc, inferred from the weighted mean of the two maser parallax measurements of the neighboring source I(N),][]{Chibueze14,Reid14} allows us to achieve a linear resolution of $\approx500$~au.  We confirm that MM1 now exhibits strong 6.7~GHz \methanol\/ maser emission.
This is the first such detection of a Class~II \methanol\/ maser from this member of the protocluster in three decades of previous interferometric observations.  We also find a new source of excited OH maser emission toward the non-thermal radio source CM2 to the north of MM1.  The locations and relative strengths of the new maser spots in relation to images of thermally excited methanol offer a unique test of maser pumping theory and provide important clues to the origin of maser emission and its response to protostellar outbursts.

\section{Observations}

\subsection{VLA}

The VLA observing parameters in the two frequency bands (C and K) are presented in Table~\ref{obs}.  In addition to the coarse resolution spectral windows used to image the continuum emission in C-band, we observed several maser transitions with high spectral resolution 2~MHz-wide correlator windows using a correlator recirculation factor of 8.  The 6.66852~GHz (hereafter 6.7~GHz) transition of \methanol\/ 5(1)-6(0) A$^+$ (\Elower49~K) was observed with a channel spacing of 1.953~kHz (0.0878~\kms) over a span of 90~\kms\/ centered on the local standard of rest (LSR) velocity of $-7$~\kms.  Three lines of excited-state OH maser emission were also observed: $J$=1/2-1/2, $F$=0-1 at 4.66024~GHz (\Elower182~K) observed with 0.0628~\kms\/ channels over a span of 64~\kms, and $J$=5/2-5/2, $F$=2-2 at 6.03075~GHz and $F$=3-3 at 6.03509~GHz (both having \Elower120~K) observed with 0.0970~\kms\/ channels over a span of 99~\kms.  The 6~GHz lines were observed using the 8-bit digitizers. The data were calibrated using the VLA pipeline\footnote{See \url{https://science.nrao.edu/facilities/vla/data-processing/pipeline/scripted-pipeline} for more information.} with some additional flagging applied (the pipeline { task that attempts to flag} radio frequency interference was not applied { to the science target in order to avoid flagging the maser features}).
The pipeline applies Hanning smoothing to the data, which reduces ringing from strong spectral features. The bright 6.7~GHz maser emission was used to iteratively self-calibrate the 6.7~GHz line emission and the resulting solutions were applied to the other line and continuum data.  After excising line emission from the broad band continuum  spectral windows, multi-term multi-scale clean was used with scales of 0, 5, and 15 times the image pixel size to produce the continuum image. The bright UCHII region (MM3 = NGC~6334F) limits the dynamic range of the continuum image causing the rms noise in the image to vary significantly as a function of position.  After subtracting the continuum in the uv-plane, spectral cubes of the maser emission were made with a channel spacing (and effective spectral resolution) of 0.15~\kms\/. Stokes I cubes were made for the relatively unpolarized 6.7~GHz transition, while right and left circular polarization (RCP and LCP) cubes were made for the strongly polarized excited OH transitions. The typical noise in a signal-free channel is 2.2~\mjb\/ for the \methanol\/ Stokes I cube, and 3.1~\mjb\/ for the RCP and LCP 6~cm excited OH cubes.  For all maser lines, although recirculation was used, the velocity extent of the emission is much smaller than the {total bandwidth} of the spectral window employed, thus none of the maser channels are affected by any spectral artifacts at high dynamic range due to phase serialization in the correlator \citep{Sault13}. 


The epoch 2011.6 K-band (1.3~cm) observations reported here employed narrow spectral windows (with limited total bandwidth) to target a range of spectral lines at high spectral resolution, including the H64$\alpha$ radio recombination line (RRL) at 24.50990~GHz that we discuss in this work.  These data were calibrated in CASA using manually generated scripts prior to the development of the pipeline. The H64$\alpha$ cube was created with a velocity resolution of 2.0~\kms\/. We estimate that the absolute position uncertainties for all of the the VLA data reported here are smaller than 50 mas.  

\subsection{ALMA}

We also present spectral line data from ALMA project 2015.A.00022.T, for which the observing parameters are described in more detail in \citet{Hunter17}. The thermal transition of methanol 11(2)-10(3) $A^-$ at 279.35191~GHz (\Elower177~K) was observed with a channel spacing of 976.5625 kHz (1.05~\kms) and a velocity resolution of 1.21 \kms.  The data cube was cleaned with a robust parameter of 0 and has a beamsize of $0\farcs29 \times 0\farcs22$ at position angle $-84^\circ$.  The rms noise achieved is 1.7 \mjb\/ in 1.1~\kms\/ wide channels.  We constructed a moment image of the peak intensity across the 22 channels spanning the entire line. We estimate that the absolute position uncertainties for the ALMA data presented here are smaller than 50 mas. 

\section{Results}

\begin{figure*}[ht!]   
\includegraphics[width=1.0\linewidth]{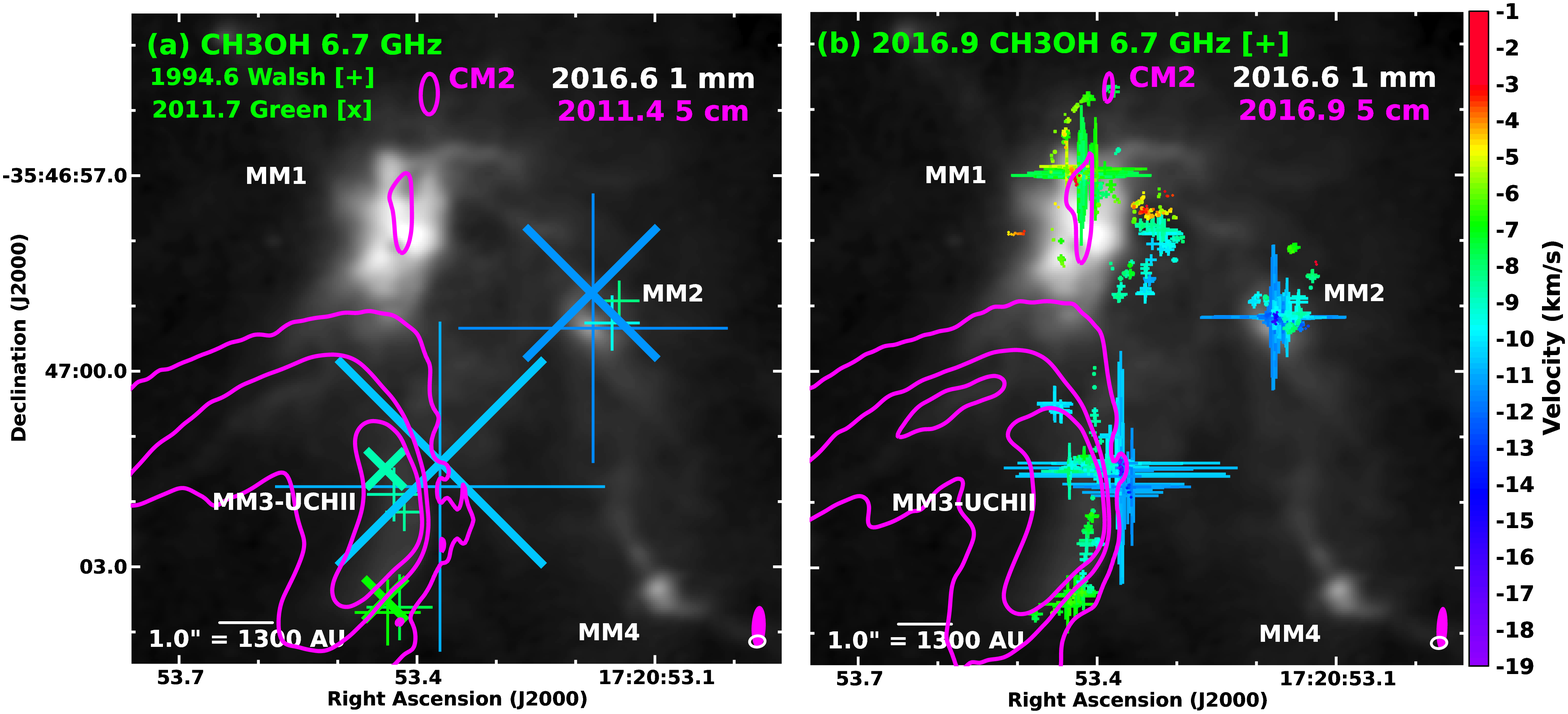}
\caption{(a) The methanol maser positions observed by the Australia Telescope Compact Array (ATCA) on July 31, 1994  \citep[][$+$ symbols, { detection threshold $\approx1$~Jy}]{Walsh98} 
and September 2, 2011 \citep[][x symbols, { detection threshold $\approx0.6$~Jy}]{Green15}
are overlaid on an epoch 2016.6 ALMA 1~mm continuum image in greyscale \citep{Hunter17}. Epoch 2011.4 VLA 5~cm continuum contours { \citep{Brogan16}} are also overlaid in magenta at levels of (4, 260, and 600) $\times$ the rms level of $3.7 \times 10^{-2}$~\mjb.  Maser intensity is indicated by the diameter of the symbol (the diameter is proportional to the square root of the intensity), while velocity is indicated by the color of the symbol.  The millimeter continuum sources are labeled for reference. (b) Same as (a) but the epoch 2016.9 5~cm continuum (magenta countours) and methanol masers ($+$ symbols) observed with the VLA are overlaid.  { The continuum contours are (4, 260, and 600) $\times$ the rms level of $2.2 \times 10^{-2}$~\mjb. The synthesized beam shape is slightly different between the epochs. }
\label{methanolimage}}
\end{figure*}


\begin{deluxetable*}{ccccccc}   
\tabletypesize{\scriptsize}
\tablewidth{0pc}
\tablecaption{Fitted properties of the 6.7~GHz methanol masers\label{methanoltable}}  
\tablecolumns{6}
\tablehead{\colhead{Number\tablenotemark{a}} & \colhead{Association} & \colhead{Velocity channel} & \multicolumn{2}{c}{Fitted Position (J2000)} & \multicolumn{1}{c}{Flux Density} \\    
& & (\kms) & \colhead{R.A.} & \colhead{Dec.} &  (Jy)\tablenotemark{b} }
\startdata
  1 & UCHII-Met4 & -18.80 & 17:20:53.3729 & -35:47:01.474 & 0.0119 (0.0015)\\
  2 & UCHII-Met4 & -18.65 & 17:20:53.3721 & -35:47:01.382 & 0.00893 (0.00148)\\
  3 & UCHII-Met4 & -17.75 & 17:20:53.3703 & -35:47:01.456 & 0.0114 (0.0015)\\
  4 & UCHII-Met4 & -17.60 & 17:20:53.3715 & -35:47:01.356 & 0.0119 (0.0015)\\
  5 & UCHII-Met4 & -17.45 & 17:20:53.3705 & -35:47:01.393 & 0.00988 (0.00147)\\
  6 & UCHII-Met4 & -17.30 & 17:20:53.3696 & -35:47:01.361 & 0.0110 (0.0015)\\
  7 & UCHII-Met4 & -17.15 & 17:20:53.3663 & -35:47:01.416 & 0.0109 (0.0015)\\
  8 & UCHII-Met4 & -17.00 & 17:20:53.3710 & -35:47:01.394 & 0.0111 (0.0015)\\
  9 & UCHII-Met4 & -16.85 & 17:20:53.3685 & -35:47:01.503 & 0.0150 (0.0014)\\
 10 & UCHII-Met4 & -16.70 & 17:20:53.3684 & -35:47:01.496 & 0.0171 (0.0015)\\
\enddata
\tablenotetext{a}{Entries are sorted by velocity}
\tablenotetext{b}{Uncertainties in the fitted value are listed in parentheses, and do not include the flux calibration uncertainty.}
\tablenotetext{}{(This table is available in its entirety in a machine-readable form in the online journal. A portion is shown here for guidance regarding
its form and content.)}
\end{deluxetable*}

\begin{figure*}[ht!]  
\includegraphics[width=0.49\linewidth]{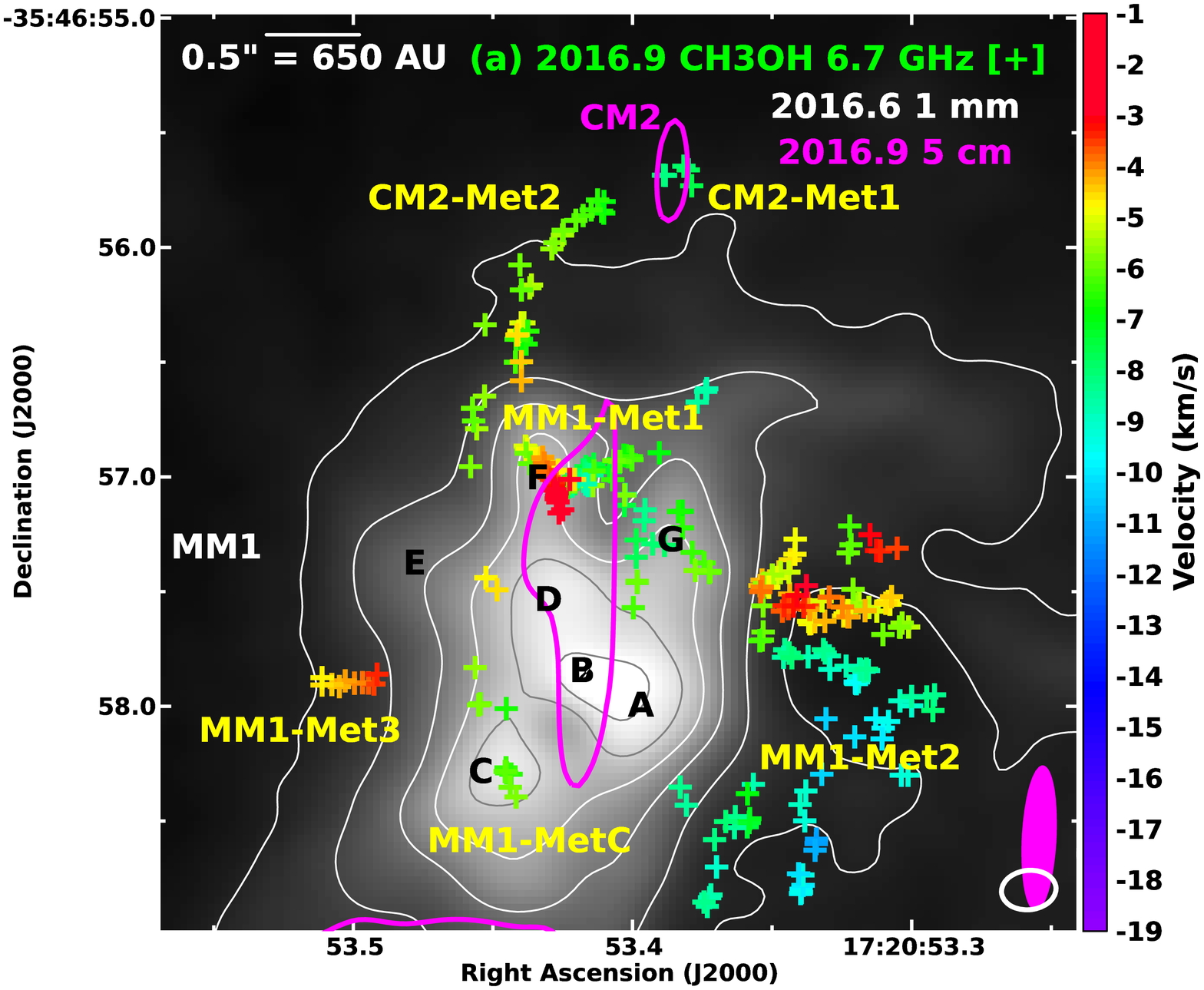}
\includegraphics[width=0.49\linewidth]{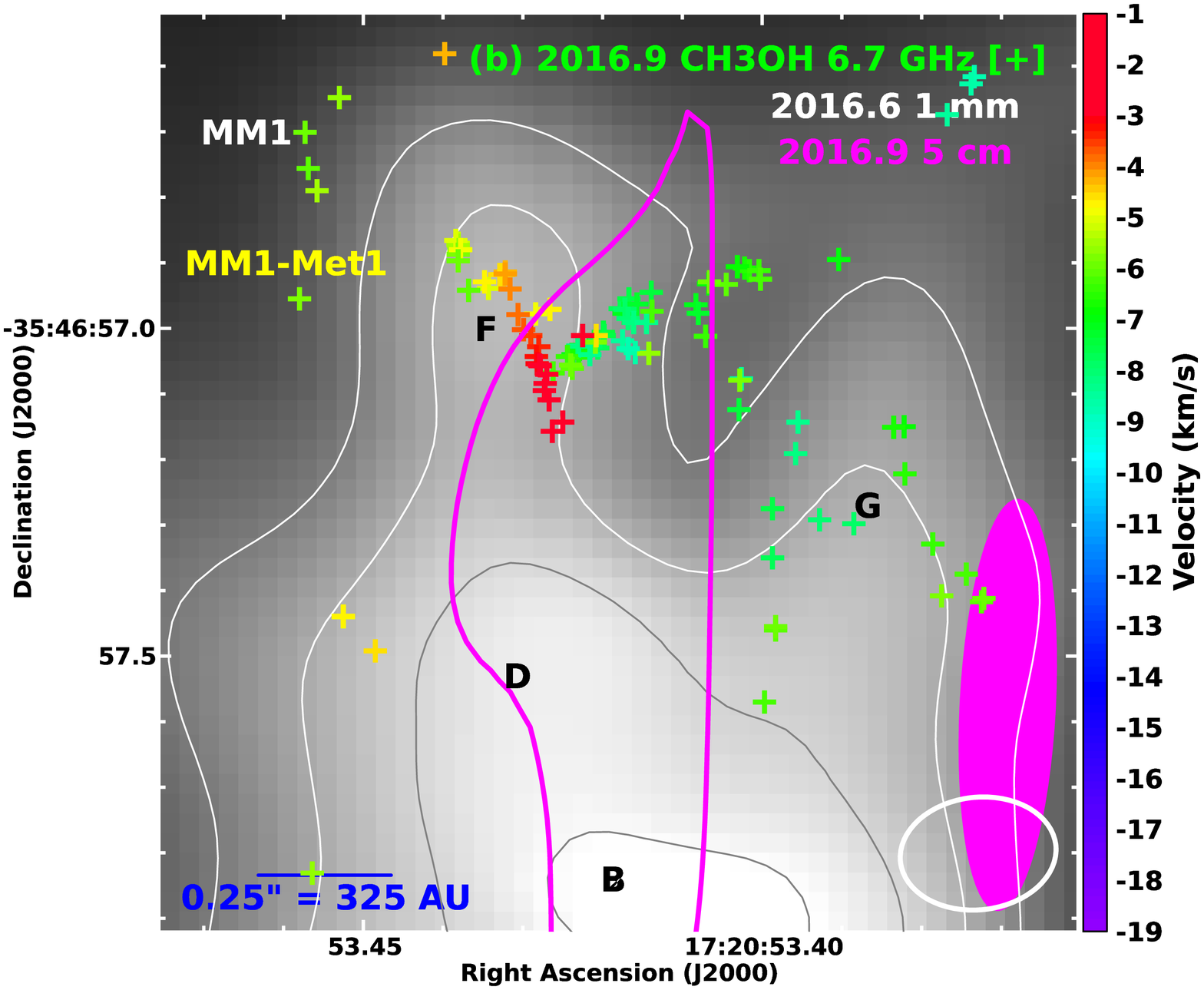}
\includegraphics[width=0.49\linewidth]{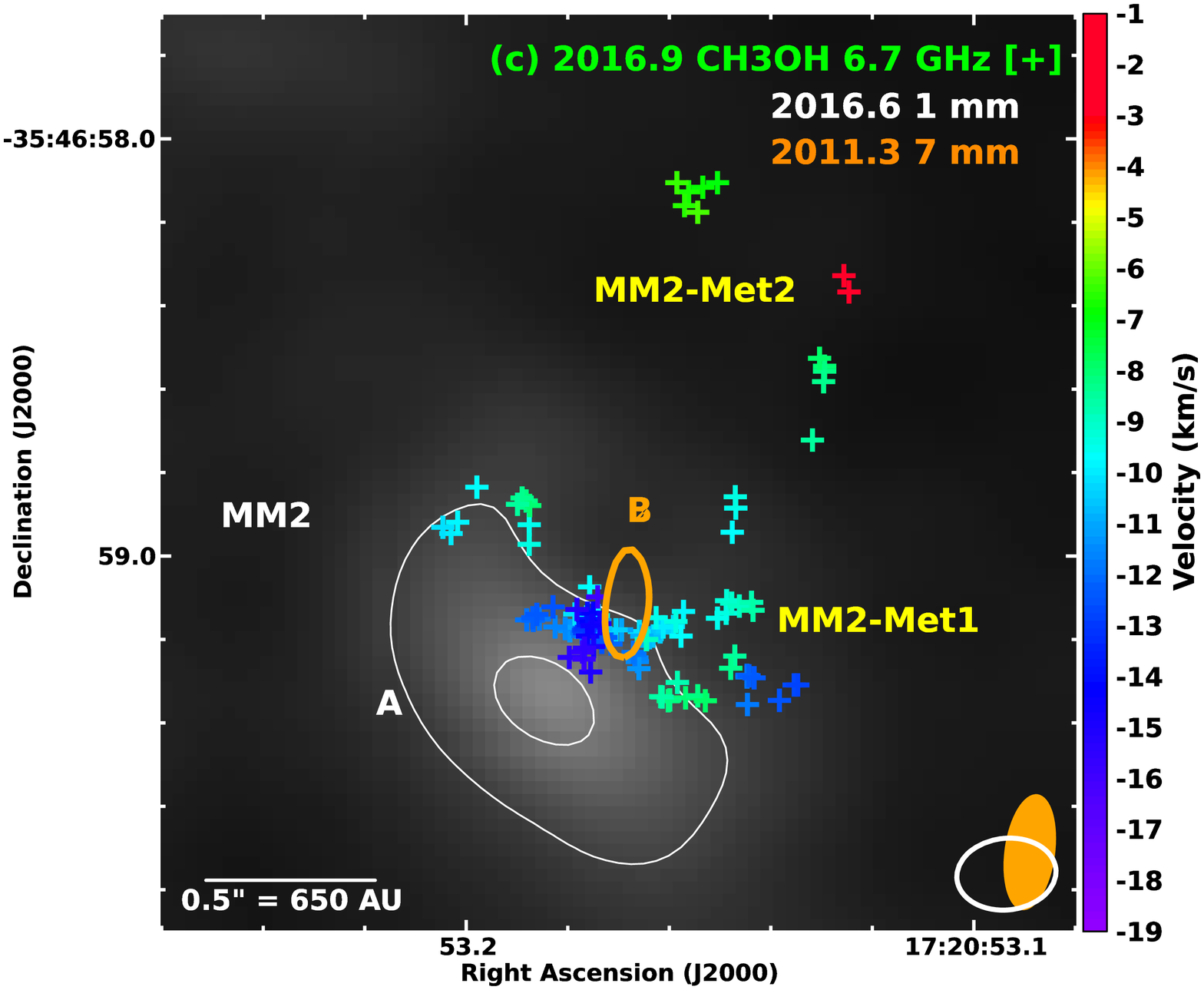}
\includegraphics[width=0.49\linewidth]{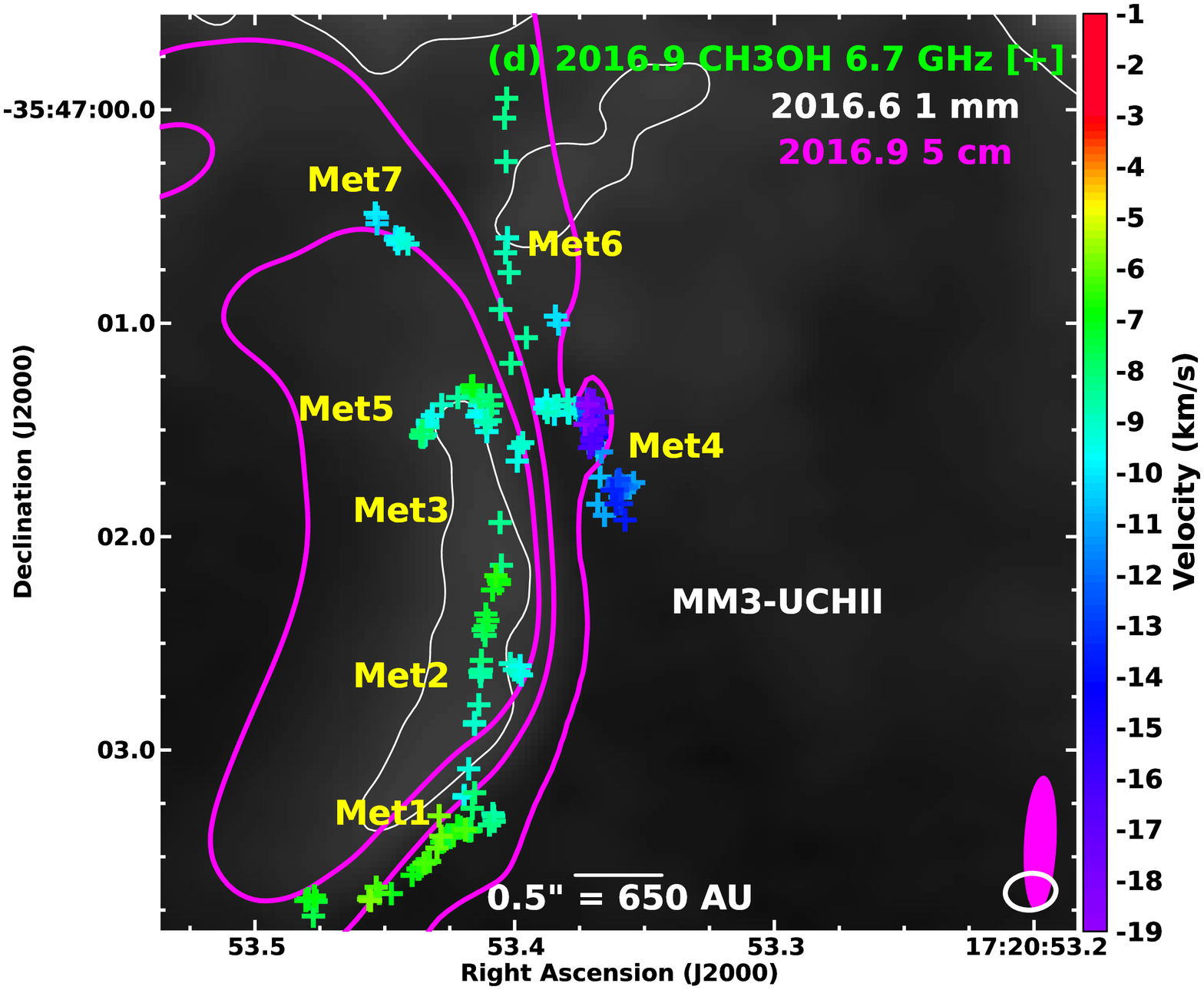}
\caption{Panels (a), (b), (c), and (d) show close up views of regions with 6.7~GHz \methanol\/ masers from Fig.~\ref{methanolimage}b; the magenta VLA 5~cm contour levels are the same. The greyscale and white (or gray) contours show the ALMA 1~mm 
dust emission, with contour levels of (a) 15, 56, 140, 245, 364, 490~\mjb\/; (b) 140, 245, 364, 490~\mjb\/, (c) 50, 140~\mjb\/; and (d) 21~\mjb\/. Panel (c) shows 
a single { 7~mm contour (in orange) denoting the intensity level 0.72~\mjb\/ for the epoch 2011.3} to indicate the position of MM2B \citep{Brogan16}. In panels (a), (b), and (c) the black, white, or orange letters denote the compact dust continuum sources defined by \citet{Brogan16}. In each panel, the maser associations from Table~\ref{methanolassoc} are labelled in yellow for reference, and the synthesized beam sizes are shown in the lower right-hand corners.  The maser locations are marked by $+$ symbols with fixed size.
\label{methanolzoom}}
\end{figure*}

\subsection{Radio continuum}

Contours from the new 5~cm continuum image are shown in Fig.~\ref{methanolimage} alongside those from the epoch 2011.4 image \citep{Brogan16}. The morphology { and intensity} of the emission are consistent given the difference in synthesized beams, and the lower sensitivity and image fidelity of the earlier data.  In particular, the compact non-thermal source CM2 reported by \citet{Brogan16} has persisted to the current epoch, as has the north-south jet-like feature centered on the grouping of millimeter sources designated MM1. The fitted position of the continuum emission from the point source CM1 located outside the field of view of Fig.~\ref{methanolimage} to the west of the central protocluster \citep{Brogan16} agrees between the 2011.4 and 2016.9 epochs to within $0\farcs03$; thus, there is no evidence for any significant instrumental shift in astrometry between the two epochs. We also note that the proper motion of the parent cloud \citep[measured by the Very Long Baseline Array,][]{Wu14} accumulated over the 5.5~years between the two epochs corresponds to only (+1.7, $-$12)~milliarcsec, which is undetectable at the current resolution. Further details about the centimeter wavelength continuum emission will be presented in a future multi-wavelength paper { (Brogan et al. 2018, in preparation)}.

\begin{deluxetable*}{ccccccc}   
\tabletypesize{\scriptsize}
\tablewidth{0pt}
\tablecaption{Summary of 6.7~GHz methanol maser associations\label{methanolassoc}}  
\tablecolumns{7}
\tablehead{\colhead{Association} & \multicolumn{2}{c}{Centroid position (J2000)} & Integrated flux & Velocity min, peak, max & Peak flux density & $T_{B}$\tablenotemark{a}\\    
& \colhead{R.A.} & \colhead{Dec.} &  (Jy \kms) & (\kms) & (Jy) & (10$^6$ K) }
\startdata
  CM2-Met1 & 17:20:53.381 & -35:46:55.68 & 2.00 & -8.00, -7.55, -7.40 & 5.03 & 0.849 \\
  CM2-Met2 & 17:20:53.413 & -35:46:55.84 & 2.72 & -6.80, -6.65, -5.30 & 4.98 & 0.840 \\
       CM2 Total & ... & ... & 4.72 & -8.00, -7.55, -5.30 & ... & ...\\
  MM1-Met1 & 17:20:53.417 & -35:46:56.98 & 741 & -8.60, -7.25, -1.55 & 545 & 92.0 \\
  MM1-Met2 & 17:20:53.327 & -35:46:58.01 & 70.8 & -11.00, -9.35, -2.75 & 55.4 & 9.35 \\
  MM1-Met3 & 17:20:53.499 & -35:46:57.90 & 0.0925 & -4.55, -3.50, -3.20 & 0.111 & 0.0188 \\
  MM1-MetC & 17:20:53.439 & -35:46:58.30 & 0.924 & -8.30, -6.20, -5.60 & 1.51 & 0.255 \\
       MM1 Total & ... & ... & 813 & -11.00, -7.25, -1.55 & ... & ...\\
  MM2-Met1 & 17:20:53.174 & -35:46:59.17 & 644 & -15.95, -11.15, -7.70 & 583 & 98.4 \\
  MM2-Met2 & 17:20:53.139 & -35:46:58.39 & 3.55 & -8.30, -7.85, -2.30 & 5.59 & 0.943 \\
       MM2 Total & ... & ... & 648 & -15.95, -11.15, -2.30 & ... & ...\\
UCHII-Met1 & 17:20:53.430 & -35:47:03.46 & 87.7 & -9.35, -6.65, -5.60 & 93.0 & 15.7 \\
UCHII-Met2 & 17:20:53.410 & -35:47:02.56 & 13.7 & -9.35, -8.00, -6.35 & 10.9 & 1.83 \\
UCHII-Met3 & 17:20:53.405 & -35:47:02.04 & 0.203 & -8.15, -8.00, -8.00 & 0.707 & 0.119 \\
UCHII-Met4 & 17:20:53.370 & -35:47:01.52 & 1760 & -18.80, -10.40, -8.15 & 1820 & 307 \\
UCHII-Met5 & 17:20:53.430 & -35:47:01.49 & 61.2 & -9.50, -8.60, -6.80 & 88.9 & 15.0 \\
UCHII-Met6 & 17:20:53.389 & -35:47:00.91 & 5.80 & -10.10, -10.10, -8.00 & 16.9 & 2.86 \\
UCHII-Met7 & 17:20:53.451 & -35:47:00.53 & 23.2 & -10.25, -10.10, -9.05 & 36.3 & 6.13 \\
     UCHII Total & ... & ... & 1950 & -18.80, -10.40, -5.60 & ... & ...\\
     Grand Total & ... & ... & 3420 & -18.80, -10.40, -1.55 & ... & ...\\
\enddata
\tablenotetext{a}{Lower limits to the brightness temperature $T_B$ using the synthesized beam; the physical sizes of the maser spots are likely to be smaller than the beam ($\sim 520$~au).}
\end{deluxetable*}

\subsection{6.7 GHz methanol masers}

Because the degree of circular polarization of the 6.7~GHz maser emission is very low, we analyzed the Stokes I cube.  
{Strong emission appears in many directions within the central protocluster. The spectrum toward any given pixel typically extends over several \kms\/ with components of rather different strength and width appearing at different velocities, and thus is not amenable to Gaussian spectral profile fitting. Instead,} we fit each channel of the cube that had significant emission ($> 6\sigma$, measured independently for each channel) with an appropriate number of {two-dimensional} Gaussian sources.  Because the spots of maser emission are likely smaller than our synthesized beam (Table~\ref{obs}), we fixed the fitted shape parameters to the synthesized beam.  We used the pixel position of each emission peak as the initial guess for the centroid position of each fit. In channels with complicated emission, we also set the intensity of each peak as an initial guess.  After manual inspection of all the fits, a few of the weakest features that formally met the sensitivity criterion could be attributed to imaging artifacts in dynamic-range limited channels and were subsequently discarded. The fitted { per-channel} positions and { per-channel} flux densities of the masers are given in Table~\ref{methanoltable}, and their positions and relative intensities are shown in Figure~\ref{methanolimage}{ b}.  For comparison, we also show previously published maser spots from two representative studies in Figure~\ref{methanolimage}a.  As this figure demonstrates, the general location and strength of the masers toward MM2 and MM3 are consistent with past observations \citep{Walsh98,Green15}. To summarize the salient maser properties in a compact format we have identified several distinct maser ``associations'' toward each of the continuum sources with detected 6.7~GHz maser emission: MM1, MM2, CM2, and the UCHII region MM3.  The assignment of each fitted spot to an association is provided in a column in Table~\ref{methanoltable}, and the properties and general locations of the associations are summarized in Table~\ref{methanolassoc} and Figure~\ref{methanolzoom}a, b, c, and d, respectively.  

Lower limits to the peak 6.7~GHz maser brightness temperatures  ($T_B$) for the associations are also given in Table~\ref{methanolassoc} derived using the peak intensity and the synthesized beam size (Table~\ref{obs}). Even given our modest resolution, lower limits as high as $T_B = 8.5\times 10^5$, $1.1\times 10^8$, $1.5\times 10^8$, and $3.1\times 10^8$~K are found for the CM2, MM1, MM2, and MM3-UCHII regions, respectively. A number of studies using very long baseline interferometry (VLBI) have explored the intrinsic size of 6.7~GHz masers, with several finding that significant flux (up to 50\%) can be missed by very high angular resolution observations, suggesting a core-halo type of morphology with a very compact core of a few au and extended halos of a few hundred au (i.e., of order the minor axis of our synthesized beam, $0\farcs2$) \citep[see for example][]{Minier02,HarveySmith06}. Therefore, we are unlikely to be missing significant flux, but it is possible that the maser spot cores have significantly higher $T_B$ than our current lower limits. 

\begin{figure}[ht!]               
\includegraphics[width=1.0\linewidth]{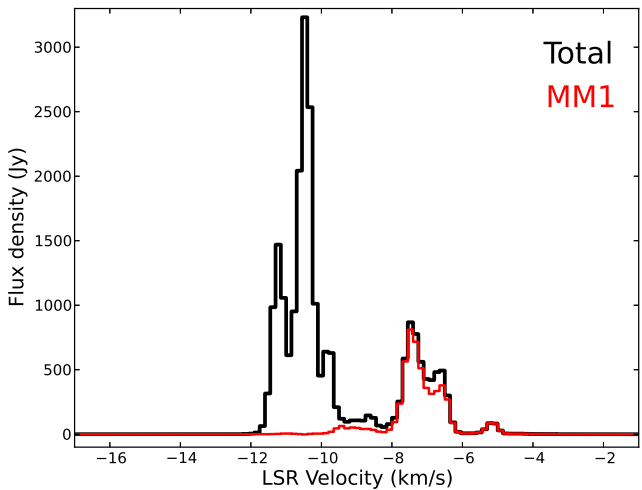}
\caption{Integrated Stokes I spectra from the epoch 2016.9 6.7~GHz methanol maser cube integrated over all the maser features (black spectrum) and over an elliptical region surrounding only the maser features associated with MM1 (red spectrum). \label{methanol_spectra}}
\end{figure}

\subsubsection{MM1 and CM2 region}
In epoch 2016.9, many 6.7~GHz \methanol\/ maser features have appeared within the dust continuum source MM1 (including the associations MM1-Met1 and MM1-MetC) and along its western periphery (MM1-Met2), eastern periphery (MM1-Met3), and northern peripheries (CM2-Met1, and CM2-Met2).  A zoom to the area around MM1 and CM2 is shown in Figure~\ref{methanolzoom}a. These are the first detections of Class~II methanol maser emission toward this region within the NGC6334I protocluster from nearly 30 years of interferometric observations, all of which were sufficiently sensitive to detect masers of this strength.  Previous epochs at 6.7~GHz include January 1992 \citep{Norris93}, May 1992 \citep{Ellingsen96,Ellingsen96etal}, July 1994 \citep{Walsh98}, July 1995 \citep{Caswell97}, September 1999 \citep{Dodson12}, March 2005 \citep{Krishnan13}, May 2011 \citep{Brogan16}, and August 2011 \citep{Green15}. In all of these epochs, detections of this maser transition in this field originate from the UCHII region (NGC6334F) and MM2. In addition, the original VLBI imaging of the Class~II 12.2~GHz methanol maser { transition} did not detect any features toward MM1 \citep{Norris88}, { nor did the May 1992 ATCA observations \citep{Ellingsen96etal}}.

The brightest 6.7~GHz features in MM1 (peak = 545~Jy at $-7.25$~\kms) reside in the valley of dust continuum emission between the 1~mm dust continuum sources MM1F and MM1G (in the MM1-Met1 association).  This area lies about 1000~au north of the hypercompact HII region MM1B, which is the proposed central driving source of the millimeter outburst \citep{Hunter17}.  The velocity of the brightest feature coincides with the LSR velocity of the thermal molecular gas \citep[-7.3~\kms,][]{Zernickel12}.  The integrated spectrum of MM1 (all associations shown in Fig.~\ref{methanolzoom}a) compared to the total field (Figure~\ref{methanol_spectra}) further demonstrates that the dominant emission from this region arises from $-6$ to $-8$~\kms. 

Within the MM1-Met1 association (see Fig.~\ref{methanolzoom}b), a line of strong spots form the western side of a ``V''-shaped structure, with a weaker line of redshifted features south of MM1F forming the eastern side. Additional strong features appear on the dust continuum source MM1F while weaker features near the LSR velocity lie around MM1G and extend northward from MM1F (including the association CM2-Met2). The northernmost spots (in association CM2-Met1) are coincident with the non-thermal radio source CM2.  Another set of near-LSR features (MM1-MetC) coincide with the millimeter source MM1C, however the three brightest and most central millimeter sources (MM1A, B, and D) are notably lacking in any maser emission.  A large number of moderate strength, primarily blueshifted masers (MM1-Met2) lie in cavities of the dust emission located west and southwest of MM1A. Finally, a number of weak, moderately redshifted features (MM1-Met3) lie eastward in an depression in the dust emission that originates between MM1C and MM1E. 

\subsubsection{MM2 region}
A zoom to the area around MM2 is shown in Figure~\ref{methanolzoom}c.  The strongest masers (MM2-Met1 in Tables~\ref{methanoltable} and \ref{methanolassoc}) are situated between the 1~mm source MM2A and the 7~mm source MM2B and are predominantly blueshifted by 5-10~\kms\/ from the LSR velocity of the thermal molecular gas.  A subgroup of masers near the LSR velocity are centered just west of the MM2B.  Weaker masers lie $\sim1''$ to the northwest (MM2-Met2 in Tables~\ref{methanoltable} and \ref{methanolassoc}) including two weak redshifted spots.  

\subsubsection{MM3 region (UCHII region)}
A zoom to the area around MM3 is shown in Figure~\ref{methanolzoom}d.  The arrangement of 6.7~GHz maser spots toward the UCHII region is strikingly filamentary, and seems to closely trace the shape of the 5~cm continuum emission, 
especially toward the southern and middle regions. 
{ These masers may arise in a layer of compressed gas just outside the UCHII region.}  
Most of these spots lie near the LSR velocity, becoming somewhat blueshifted in the northern half.  The strongest spots (Met4) are significantly blueshifted (by up to 12~\kms) and cluster around a small knot in the continuum emission located about $1''$ northwest of the 5~cm UCHII region peak.

\begin{figure*}[ht!]               
\includegraphics[width=0.96\linewidth]{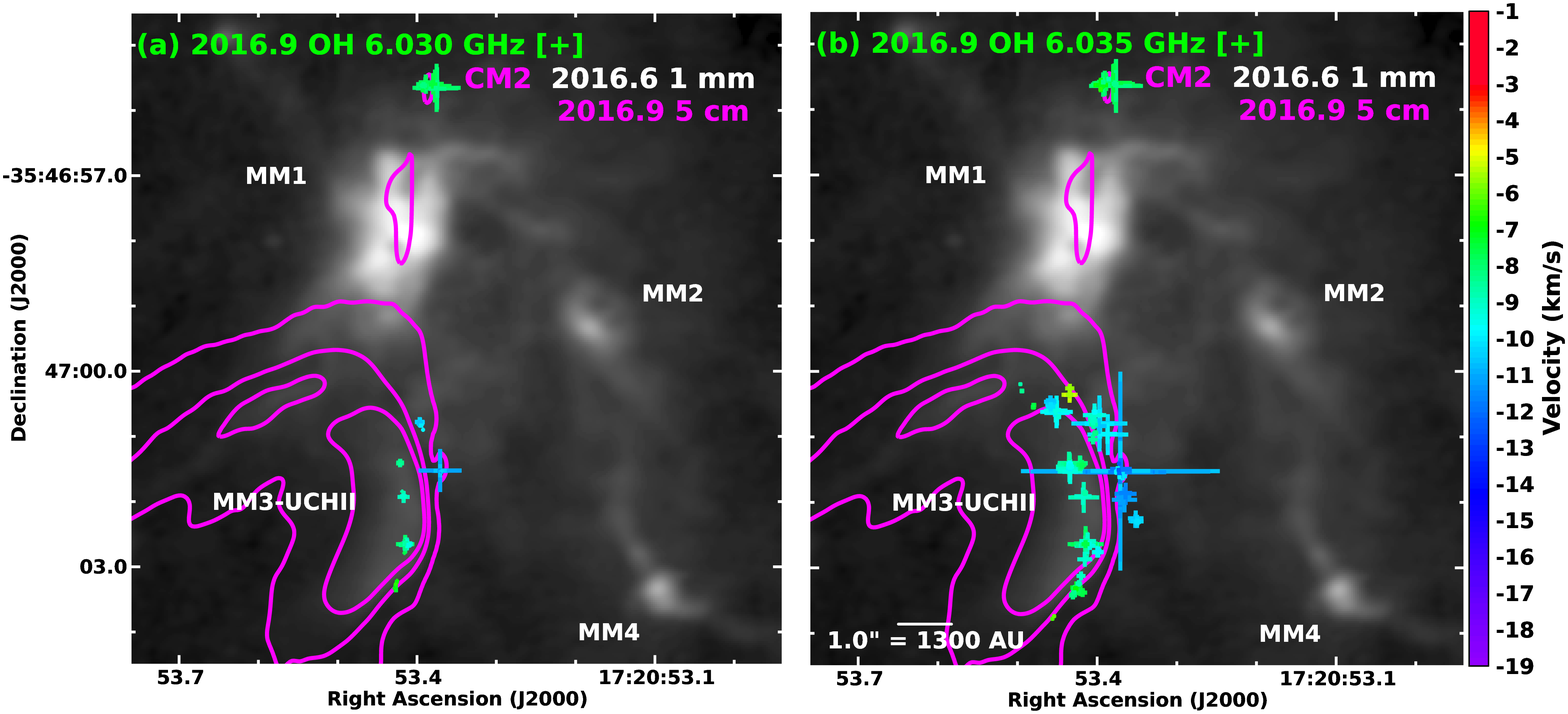}
\caption{Excited-state OH masers: (a) The fitted positions of the 6.030~GHz transition ($+$ symbols) are overlaid on an epoch 2016.6 ALMA 1 mm continuum image in greyscale \citep{Hunter17}. In both panels, the VLA 5 cm continuum contours from epoch 2016.9 are overlaid in magenta with contour levels the same as Fig.~\ref{methanolimage}b. As in Fig.~\ref{methanolimage}, the maser intensity is indicated by the size of the symbol, while velocity is indicated by the color.  Continuum sources are labeled for reference;  (b) Same as (a) but for the 6.035~GHz transition.
\label{ohmasers}}
\end{figure*} 

\begin{deluxetable*}{ccccccc}          
\tabletypesize{\scriptsize}
\tablewidth{0pc}
\tablecaption{Fitted properties of the OH 6.030~GHz masers (left circular polarization) \label{oh6030LLtable}}  
\tablecolumns{6}
\tablehead{\colhead{Number\tablenotemark{a}} &  \colhead{Association} & \colhead{Velocity channel} & \multicolumn{2}{c}{Fitted Position (J2000)} & \multicolumn{1}{c}{Flux Density} \\    
& & (\kms) & \colhead{R.A.} & \colhead{Dec.} &  (Jy)\tablenotemark{b}}
\startdata
  1 & UCHII-OH4 & -11.00 & 17:20:53.3718 & -35:47:01.573 & 0.0241 (0.0025)\\
  2 & UCHII-OH4 & -10.85 & 17:20:53.3722 & -35:47:01.573 & 0.0903 (0.0022)\\
  3 & UCHII-OH6 & -10.85 & 17:20:53.3975 & -35:47:00.737 & 0.0294 (0.0022)\\
  4 & UCHII-OH4 & -10.70 & 17:20:53.3712 & -35:47:01.531 & 0.326 (0.002)\\
  5 & UCHII-OH6 & -10.70 & 17:20:53.3967 & -35:47:00.768 & 0.203 (0.002)\\
  6 & UCHII-OH4 & -10.55 & 17:20:53.3709 & -35:47:01.525 & 2.84 (0.01)\\
  7 & UCHII-OH6 & -10.55 & 17:20:53.3972 & -35:47:00.775 & 0.348 (0.004)\\
  8 & UCHII-OH4 & -10.40 & 17:20:53.3709 & -35:47:01.525 & 13.3 (0.1)\\
  9 & UCHII-OH6 & -10.40 & 17:20:53.3979 & -35:47:00.780 & 0.253 (0.007)\\
 10 & UCHII-OH4 & -10.25 & 17:20:53.3709 & -35:47:01.524 & 16.6 (0.1)\\
\enddata
\tablenotetext{a}{Entries are sorted by velocity}
\tablenotetext{b}{Uncertainties in the fitted value are listed in parentheses, and do not include the flux calibration uncertainty.}
\tablenotetext{}{(This table is available in its entirety in a machine-readable form in the online journal. A portion is shown here for guidance regarding
its form and content.)}
\end{deluxetable*}

\begin{deluxetable*}{ccccccc}      
\tabletypesize{\scriptsize}
\tablewidth{0pc}
\tablecaption{Fitted properties of the OH 6.030~GHz masers (right circular polarization) \label{oh6030RRtable}}  
\tablecolumns{6}
\tablehead{\colhead{Number\tablenotemark{a}} &  \colhead{Association} & \colhead{Velocity channel} & \multicolumn{2}{c}{Fitted Position (J2000)} & \multicolumn{1}{c}{Flux Density} \\    
& & (\kms) & \colhead{R.A.} & \colhead{Dec.} &  (Jy)\tablenotemark{b}}
\startdata
  1 & UCHII-OH4 & -11.30 & 17:20:53.3687 & -35:47:01.536 & 0.0305 (0.0021)\\
  2 & UCHII-OH4 & -11.15 & 17:20:53.3704 & -35:47:01.533 & 0.113 (0.002)\\
  3 & UCHII-OH4 & -11.00 & 17:20:53.3709 & -35:47:01.527 & 0.973 (0.003)\\
  4 & UCHII-OH4 & -10.85 & 17:20:53.3709 & -35:47:01.525 & 6.56 (0.01)\\
  5 & UCHII-OH4 & -10.70 & 17:20:53.3709 & -35:47:01.524 & 13.0 (0.1)\\
  6 & UCHII-OH4 & -10.55 & 17:20:53.3709 & -35:47:01.523 & 6.78 (0.01)\\
  7 & UCHII-OH4 & -10.40 & 17:20:53.3708 & -35:47:01.527 & 0.639 (0.003)\\
  8 & UCHII-OH6 & -10.40 & 17:20:53.3964 & -35:47:00.796 & 0.128 (0.003)\\
  9 & UCHII-OH4 & -10.25 & 17:20:53.3702 & -35:47:01.547 & 0.0517 (0.0027)\\
 10 & UCHII-OH6 & -10.25 & 17:20:53.3967 & -35:47:00.776 & 0.332 (0.003)\\
\enddata
\tablenotetext{a}{Entries are sorted by velocity}
\tablenotetext{b}{Uncertainties in the fitted value are listed in parentheses, and do not include the flux calibration uncertainty.}
\tablenotetext{}{(This table is available in its entirety in a machine-readable form in the online journal. A portion is shown here for guidance regarding
its form and content.)}
\end{deluxetable*}

\begin{deluxetable*}{ccccccc}   
\tabletypesize{\scriptsize}
\tablewidth{0pc}
\tablecaption{Fitted properties of the OH 6.035~GHz masers (left circular polarization) \label{oh6035LLtable}}  
\tablecolumns{6}
\tablehead{\colhead{Number\tablenotemark{a}} &  \colhead{Association} & \colhead{Velocity channel} & \multicolumn{2}{c}{Fitted Position (J2000)} & \multicolumn{1}{c}{Flux Density} \\    
& & (\kms) & \colhead{R.A.} & \colhead{Dec.} &  (Jy)\tablenotemark{b}}
\startdata
  1 & UCHII-OH4 & -12.05 & 17:20:53.3719 & -35:47:01.507 & 0.0231 (0.0022)\\
  2 & UCHII-OH4 & -11.90 & 17:20:53.3709 & -35:47:01.555 & 0.0281 (0.0024)\\
  3 & UCHII-OH4 & -11.75 & 17:20:53.3717 & -35:47:01.551 & 0.0376 (0.0024)\\
  4 & UCHII-OH4 & -11.60 & 17:20:53.3718 & -35:47:01.558 & 0.0672 (0.0021)\\
  5 & UCHII-OH4 & -11.45 & 17:20:53.3723 & -35:47:01.560 & 0.0972 (0.0025)\\
  6 & UCHII-OH4 & -11.15 & 17:20:53.3638 & -35:47:01.863 & 0.895 (0.009)\\
  7 & UCHII-OH4 & -11.15 & 17:20:53.3712 & -35:47:01.544 & 5.35 (0.01)\\
  8 & UCHII-OH4 & -11.00 & 17:20:53.3631 & -35:47:01.858 & 1.06 (0.01)\\
  9 & UCHII-OH4 & -11.00 & 17:20:53.3712 & -35:47:01.543 & 30.0 (0.1)\\
 10 & UCHII-OH4 & -10.85 & 17:20:53.3712 & -35:47:01.542 & 36.9 (0.1)\\
\enddata
\tablenotetext{a}{Entries are sorted by velocity}
\tablenotetext{b}{Uncertainties in the fitted value are listed in parentheses, and do not include the flux calibration uncertainty.}
\tablenotetext{}{(This table is available in its entirety in a machine-readable form in the online journal. A portion is shown here for guidance regarding its form and content.)}
\end{deluxetable*}

\begin{deluxetable*}{ccccccc}    
\tabletypesize{\scriptsize}
\tablewidth{0pc}
\tablecaption{Fitted properties of the OH 6.035~GHz masers (right circular polarization) \label{oh6035RRtable}}  
\tablecolumns{6}
\tablehead{\colhead{Number\tablenotemark{a}} &  \colhead{Association} & \colhead{Velocity channel} & \multicolumn{2}{c}{Fitted Position (J2000)} & \multicolumn{1}{c}{Flux Density} \\    
& & (\kms) & \colhead{R.A.} & \colhead{Dec.} &  (Jy)\tablenotemark{b}}
\startdata
  1 & UCHII-OH4 & -12.05 & 17:20:53.3731 & -35:47:01.504 & 0.0391 (0.0023)\\
  2 & UCHII-OH4 & -11.90 & 17:20:53.3723 & -35:47:01.565 & 0.0624 (0.0025)\\
  3 & UCHII-OH4 & -11.75 & 17:20:53.3718 & -35:47:01.597 & 0.0718 (0.0025)\\
  4 & UCHII-OH4 & -11.45 & 17:20:53.3644 & -35:47:01.856 & 0.617 (0.004)\\
  5 & UCHII-OH4 & -11.45 & 17:20:53.3713 & -35:47:01.539 & 2.03 (0.01)\\
  6 & UCHII-OH4 & -11.30 & 17:20:53.3643 & -35:47:01.870 & 1.78 (0.01)\\
  7 & UCHII-OH4 & -11.30 & 17:20:53.3712 & -35:47:01.541 & 17.5 (0.1)\\
  8 & UCHII-OH4 & -11.15 & 17:20:53.3712 & -35:47:01.540 & 31.2 (0.1)\\
  9 & UCHII-OH4 & -11.15 & 17:20:53.3662 & -35:47:01.899 & 0.773 (0.018)\\
 10 & UCHII-OH4 & -11.00 & 17:20:53.3671 & -35:47:01.929 & 0.727 (0.007)\\
\enddata
\tablenotetext{a}{Entries are sorted by velocity}
\tablenotetext{b}{Uncertainties in the fitted value are listed in parentheses, and do not include the flux calibration uncertainty.}
\tablenotetext{}{(This table is available in its entirety in a machine-readable form in the online journal. A portion is shown here for guidance regarding its form and content.)}
\end{deluxetable*}

\subsection{Excited-state OH 6.030 and 6.035 GHz transitions}

\begin{figure*}[ht!]  
\includegraphics[width=0.96\linewidth]{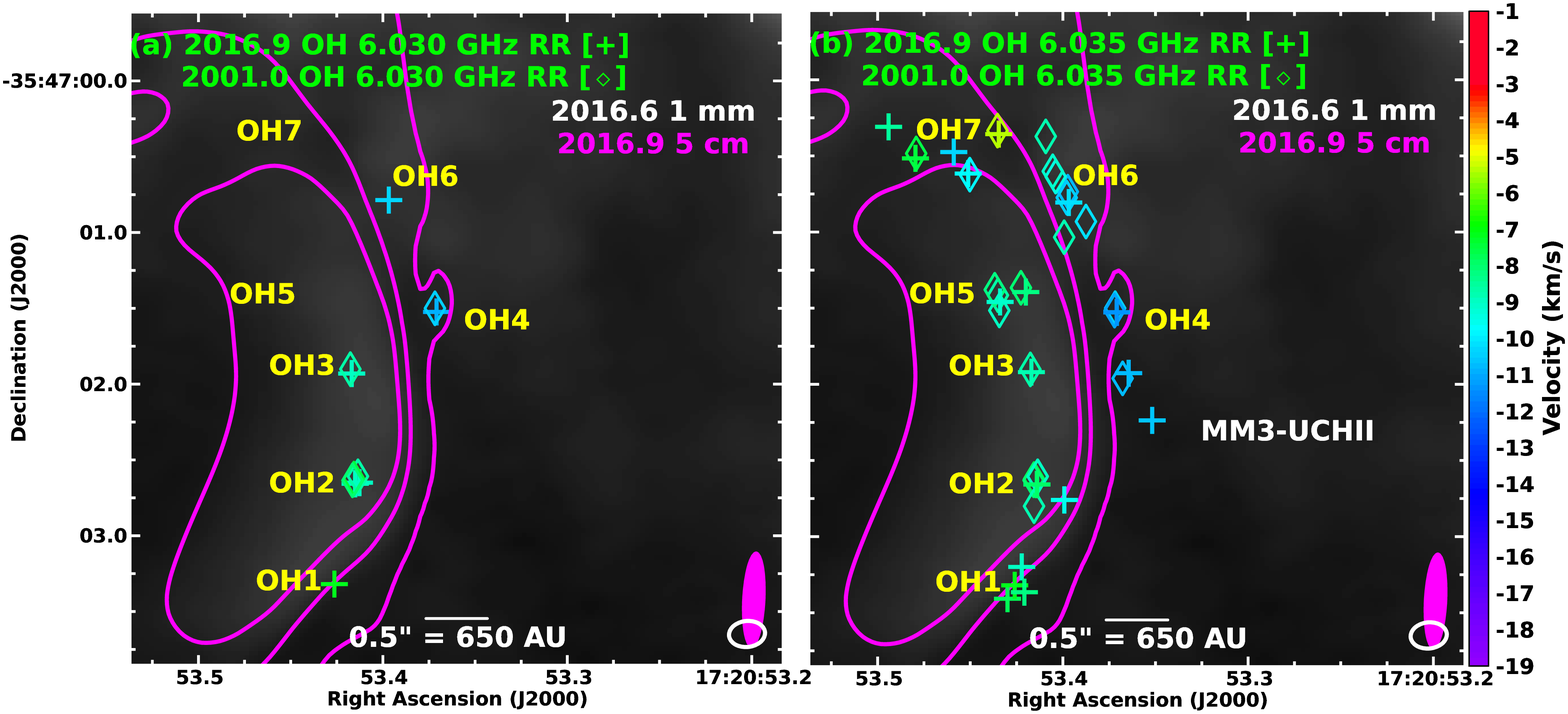}
\caption{The positions of the epoch 2016.9 VLA Zeeman pairs toward the MM3-UCHII region are plotted as crosses.  The epoch 2001.0 pairs found by \citet{Caswell11} are plotted as diamonds, after applying a -0\farcs1 shift in right ascension. The background image and contours are the same as Fig.~\ref{methanolimage}.
\label{ohcompare}}
\end{figure*}

\ngci\/ is one of the strongest known sources of the 5~cm excited-state OH masers \citep{Caswell95,Zuckerman72,Gardner70}.  {Although the excited OH spectra are generally simpler than those of the 6.7~GHz methanol, in order to match the methanol maser fitting procedure, we also fit the excited OH maser properties using the channel by channel approach.} Because the emission in these OH transitions is strongly circularly polarized, we fit the image cubes of RCP and LCP emission independently.  The fitted positions and flux densities of the two transitions (6.030 and 6.035~GHz) are given in Tables~\ref{oh6030LLtable}, \ref{oh6030RRtable}, \ref{oh6035LLtable} and \ref{oh6035RRtable}. The maser spots are shown in Figure~\ref{ohmasers}.  { Note we have not 
attempted to correct the flux densities for beam squint, but this effect will be small ($\sim 1.2\%$) toward the inner parts of the $7\arcmin$ primary beam where the masers are located (beam squint does not affect the Zeeman frequency splitting). } Since all of the excited OH masers are found in the same general areas as the 6.7~GHz methanol detections, the labels for the OH associations have the same numerical index for a given region, for example UCHII-Met1 has an analogous excited OH maser association UCHII-OH1. The spatial distribution of the two OH transitions is similar, with 6.035~GHz being significantly stronger in most locations and showing  additional associations (OH5 and OH7) toward the northern portion of the UCHII region (Figure~\ref{ohcompare}). The positions of the seven associations UCHII-OH1 through UCHII-OH7 are closely matched to those of the 6.7~GHz maser associations UCHII-Met1 through UCHII-Met7.

\begin{figure*}[ht!]                          
\includegraphics[width=0.49\linewidth]{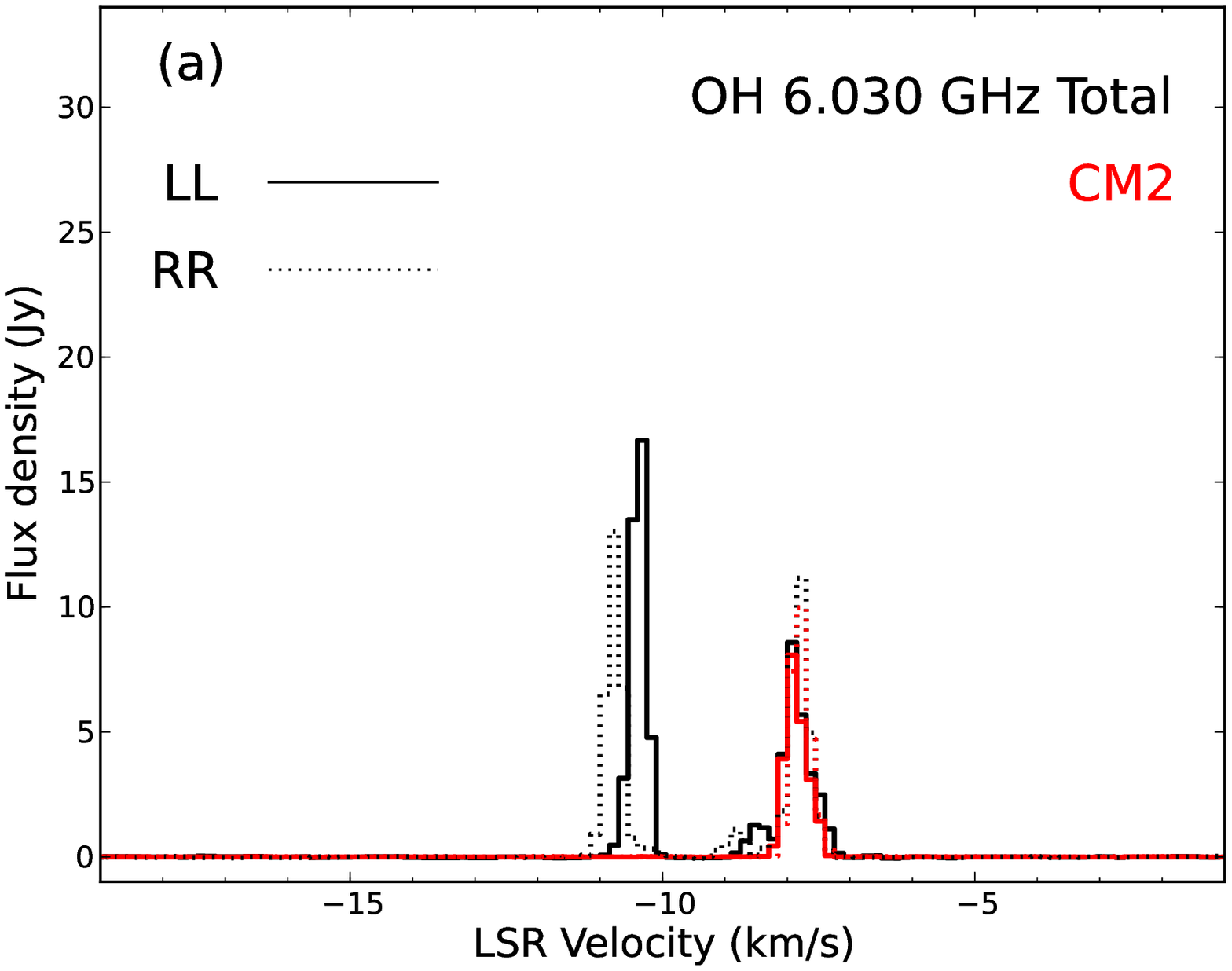}
\includegraphics[width=0.49\linewidth]{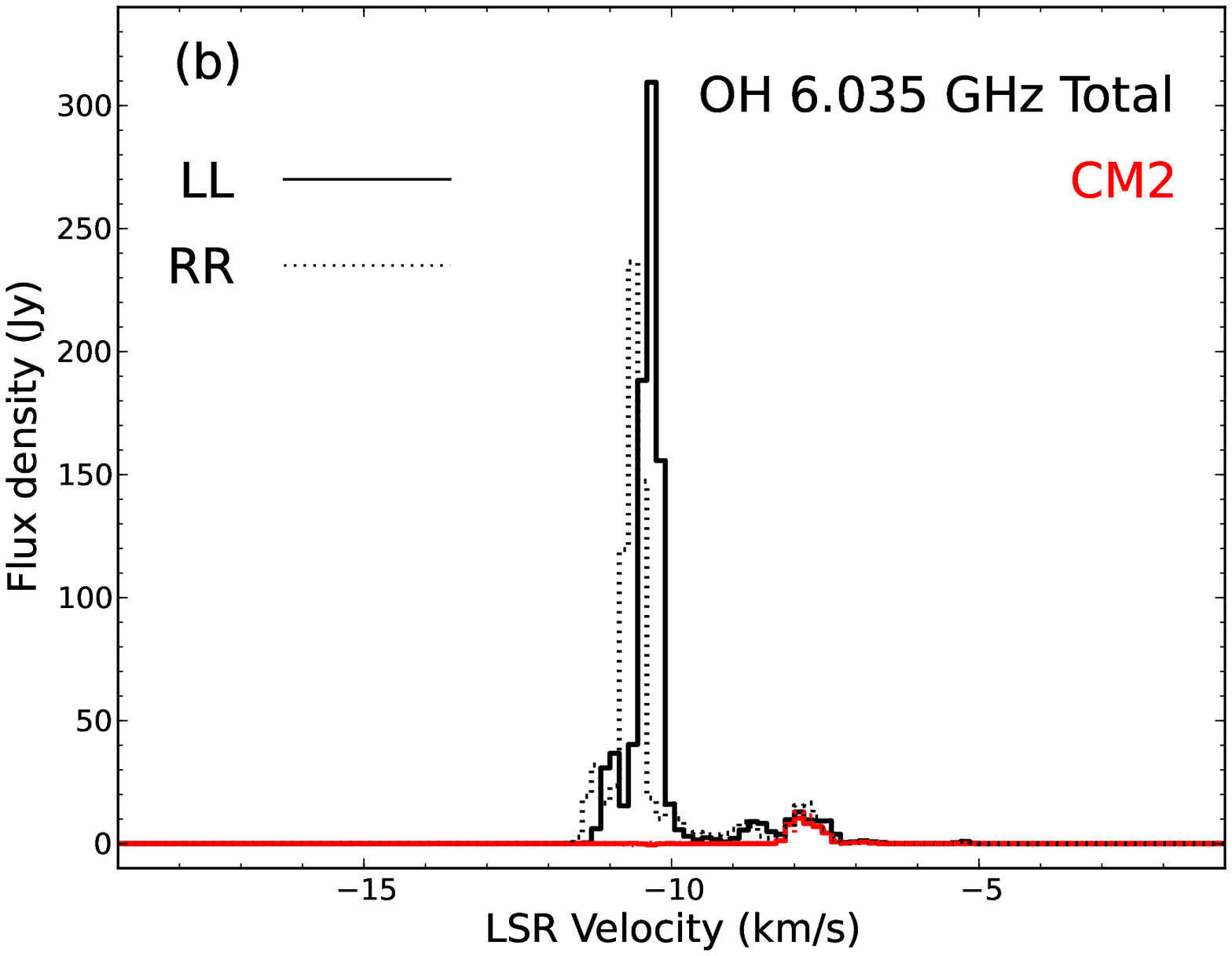}
\caption{Integrated spectra per polarization of the excited-state OH masers constructed by summing the flux density from all directions (black spectra) and from CM2 only (red spectra): (a) 6.030~GHz line; (b) 6.035~GHz line.  Note the factor of 10 change in the y-axis scale between the panels.
\label{oh_spectra}}
\end{figure*}

Integrated spectra of the cubes are shown in Figure~\ref{oh_spectra}.   These are similar to previous single-dish spectra toward this region \citep{Caswell03,Avison16}, with the exception of the strong new association toward CM2, which now dominates the emission near the LSR velocity, peaking at $-7.7$~\kms.
Both lines were also previously observed interferometrically \citep{Knowles73,Caswell97,Green15}, with the highest resolution images produced with the Long Baseline Array (LBA) in January 2001 \citep{Caswell11}, which showed the ridge of spots along the UCHII region, but no feature toward CM2 has ever been reported.  We thus conclude that the CM2 feature is associated with the millimeter outburst.  It is notable that the flux ratio of the 6.035/6.030 lines toward CM2 is abnormally low ($1.2-1.8$) compared to the rest of the spots, which have a median ratio of 23, similar to the historical ratio of 25 noted by \citet{Knowles76}. At the current resolution, a lower limit to the peak brightness temperature of the 6.030 and 6.035~GHz masers is $3\times 10^6$~K and $5\times 10^7$~K, respectively (UCHII-OH4-1b, see \S~\ref{Zeemansplitting}). From LBA observations toward NGC6334I, \citet{Caswell11} found the excited OH spot sizes to be smaller than 20~mas for the UCHII region masers, and intensities within a factor of 2 of those reported here.  From much higher angular resolution European VLBI Network (EVN) observations of the same excited OH transitions in W3(OH), \citet{Fish07} find typical maser spot sizes of order 5~mas or smaller, { which would translate to 7.5~mas or smaller at the nearer distance of NGC6334I}. Thus, the true OH brightness temperatures are likely to be orders of magnitude higher than reported here.

\subsubsection{Zeeman splitting}
\label{Zeemansplitting}
\setlength{\tabcolsep}{0.09cm}
\begin{deluxetable*}{lccccccccc} 
\tabletypesize{\scriptsize}
\tablecaption{Line of sight magnetic field ($B_{los}$) derived from OH 6.035 and 6.030 GHz transitions \label{blos}}
\tablecolumns{10}
\tablehead{\colhead{Assoc.} & \multicolumn{2}{c}{Mean Position (J2000)\tablenotemark{a}} 
& \multicolumn{3}{c}{Fitted Properties RCP}  & \multicolumn{3}{c}{Fitted Properties LCP} &  \\
& \colhead{R.A.} & \colhead{Dec.} & \colhead{Flux Density\tablenotemark{b}} & \colhead{$T_{B}$\tablenotemark{c}} & \colhead{$V_{center}$\tablenotemark{b}} 
& \colhead{Flux Density\tablenotemark{b}} & \colhead{$T_{B}$\tablenotemark{c}} & \colhead{$V_{center}$\tablenotemark{b}} 
& \colhead{$B_{los}$\tablenotemark{b}} \\
& & & \colhead{(Jy)} & \colhead{$10^6$~(K)} & \colhead{(\kms\/)} & \colhead{(Jy)} & \colhead{$10^6$~(K)} & \colhead{(\kms\/)} & \colhead{(mG)}
} 
\startdata
& & & & & OH 6.035 GHz \\
\hline
UCHII-OH1-1 & 17:20:53.422 &  -35:47:03.20 & 0.67 (0.03) & 0.11 & -8.762 (0.008) & 0.63 (0.03) & 0.11 & -8.520 (0.010) & -4.32 (0.24) \\
UCHII-OH1-2 & 17:20:53.430 &  -35:47:03.41 & 0.13 (0.02) & 0.022 & -7.533 (0.040) & 0.16 (0.01) & 0.027 & -7.391 (0.027) & -2.54 (0.86) \\
UCHII-OH1-3 & 17:20:53.421 &  -35:47:03.37 & 0.16 (0.02) & 0.027 & -8.061 (0.038) & 0.21 (0.02) & 0.036 & -7.888 (0.033) & -3.08 (0.90) \\
UCHII-OH1-4 & 17:20:53.426 &  -35:47:03.32 & 0.59 (0.02) & 0.10 & -6.923 (0.007) & 0.55 (0.02) & 0.094 & -6.644 (0.008) & -4.98 (0.19) \\
UCHII-OH2-1a & 17:20:53.399 &  -35:47:02.76 & 0.49 (0.03) & 0.083 & -9.441 (0.010) & 0.55 (0.04) & 0.094 & -9.308 (0.010) & -2.38 (0.25) \\
UCHII-OH2-1b & &                              & 0.52 (0.02) & 0.089 & -8.752 (0.013) & 0.45 (0.03) & 0.077 & -8.531 (0.018) & -3.94 (0.40) \\
UCHII-OH2-1c & &                              & 0.75 (0.03) & 0.13 & -7.692 (0.007) & 0.75 (0.04) & 0.13 & -7.386 (0.008) & -5.46 (0.18) \\
UCHII-OH2-2a & 17:20:53.414 &  -35:47:02.66 & 2.58 (0.04) & 0.44 & -8.734 (0.004) & 2.20 (0.02) & 0.37 & -8.522 (0.003) & -3.78 (0.08) \\
UCHII-OH2-2b & &                               & 1.52 (0.07) & 0.26 & -8.078 (0.005) & 1.41 (0.03) & 0.24 & -7.831 (0.003) & -4.40 (0.11) \\
UCHII-OH2-2c & &                               & 4.92 (0.05) & 0.84 & -7.671 (0.002) & 4.78 (0.03) & 0.81 & -7.375 (0.001) & -5.30 (0.04) \\
UCHII-OH3-1 & 17:20:53.417 &  -35:47:01.92 & 3.78 (0.07) & 0.64 & -8.713 (0.003) & 3.98 (0.05) & 0.68 & -8.523 (0.002) & -3.40 (0.06) \\
UCHII-OH4-1a & 17:20:53.371 &  -35:47:01.53 & 30.76 (0.95) & 5.24 & -11.156 (0.005) & 39.11 (1.09) & 6.7 & -10.899 (0.004) & -4.60 (0.12) \\
UCHII-OH4-1b & &                             & 236.89 (0.96) & 40 & -10.539 (0.001) & 298.65 (1.08) & 51 & -10.266 (0.001) & -4.89 (0.02) \\
UCHII-OH4-2a & 17:20:53.365 &  -35:47:01.93 & 11.07 (0.33) & 1.9 & -11.158 (0.006) & 13.61 (0.38) & 2.3 & -10.901 (0.005) & -4.58 (0.13) \\
UCHII-OH4-2b & &                              & 78.61 (0.36) & 13 & -10.542 (0.001) & 98.51 (0.39) & 17 & -10.268 (0.001) & -4.90 (0.02) \\
UCHII-OH4-3 & 17:20:53.352 &  -35:47:02.24 & 1.52 (0.03) & 0.26 & -10.490 (0.006) & 2.54 (0.03) & 0.43 & -10.246 (0.003) & -4.36 (0.13) \\
UCHII-OH5-1 & 17:20:53.434 &  -35:47:01.46 & 2.29 (0.13) & 0.39 & -8.913 (0.008) & 3.46 (0.04) & 0.59 & -8.645 (0.004) & -4.79 (0.16) \\
UCHII-OH5-2 & 17:20:53.420 &  -35:47:01.39 & 0.90 (0.03) & 0.15 & -8.059 (0.008) & 0.55 (0.02) & 0.094 & -7.924 (0.010) & -2.41 (0.23) \\
UCHII-OH6-1 & 17:20:53.397 &  -35:47:00.81 & 12.51 (0.17) & 2.1 & -9.993 (0.002) & 11.25 (0.14) & 1.9 & -10.217 (0.002) & +3.99 (0.06) \\
UCHII-OH6-3 & 17:20:53.404 &  -35:47:00.68 & 1.19 (0.09) & 0.20 & -9.212 (0.010) & 0.93 (0.09) & 0.16 & -9.311 (0.012) & +1.77 (0.29) \\
UCHII-OH7-1 & 17:20:53.459 &  -35:47:00.48 & 0.53 (0.01) & 0.090 & -10.200 (0.005) & 0.45 (0.01) & 0.077 & -10.463 (0.004) & +4.70 (0.11) \\
UCHII-OH7-2 & 17:20:53.451 &  -35:47:00.62 & 3.74 (0.05) & 0.64 & -9.507 (0.002) & 3.65 (0.04) & 0.62 & -9.768 (0.002) & +4.66 (0.05) \\
UCHII-OH7-3 & 17:20:53.494 &  -35:47:00.31 & 0.08 (0.01) & 0.014 & -8.281 (0.013) & 0.09 (0.01) & 0.015 & -8.071 (0.013) & -3.76 (0.33) \\
UCHII-OH7-4 & 17:20:53.480 &  -35:47:00.52 & 0.10 (0.01) & 0.017 & -7.227 (0.013) & 0.09 (0.01) & 0.015 & -7.416 (0.012) & +3.38 (0.31) \\
UCHII-OH7-5 & 17:20:53.435 &  -35:47:00.36 & 0.98 (0.01) & 0.17 & -5.117 (0.001) & 0.93 (0.01) & 0.16 & -5.175 (0.001) & +1.03 (0.03) \\
CM2-OH1-1 & 17:20:53.392 &  -35:46:55.61 & 4.94 (0.09) & 0.84 & -7.705 (0.005) & 4.63 (0.07) & 0.79 & -7.731 (0.004) & +0.47 (0.12) \\
CM2-OH1-2 & 17:20:53.378 &  -35:46:55.62 & 10.54 (0.20) & 1.8 & -7.722 (0.005) & 8.33 (0.17) & 1.4 & -7.758 (0.006) & +0.65 (0.14) \\
CM2-OH1-3 & 17:20:53.395 &  -35:46:55.63 & 0.76 (0.06) & 0.13 & -6.641 (0.014) & 0.71 (0.05) & 0.12 & -6.849 (0.011) & +3.71 (0.31) \\
\hline
& & & & & OH 6.030 GHz \\
\hline
UCHII-OH1-4 & 17:20:53.426 &  -35:47:03.32 & 0.08 (0.01) & 0.13 & -6.994 (0.012) & 0.09 (0.01) & 0.015 & -6.556 (0.010) & -5.54 (0.20) \\
UCHII-OH2-1b & 17:20:53.413 &  -35:47:02.65 & 0.57 (0.01) & 0.097 & -8.780 (0.004) & 0.66 (0.01) & 0.11 & -8.407 (0.004) & -4.71 (0.07) \\
UCHII-OH2-2a & 17:20:53.415 &  -35:47:02.66 & 0.58 (0.01) & 0.099 & -8.774 (0.004) & 0.68 (0.01) & 0.12 & -8.406 (0.003) & -4.66 (0.07) \\
UCHII-OH2-2b & &                              & 0.54 (0.01) & 0.092 & -8.181 (0.004) & 0.54 (0.01) & 0.092 & -7.811 (0.003) & -4.68 (0.06) \\
UCHII-OH2-2c & &                              & 1.35 (0.01) & 0.23 & -7.750 (0.002) & 1.33 (0.01) & 0.23 & -7.327 (0.001) & -5.35 (0.03) \\
UCHII-OH3-1 & 17:20:53.417 &  -35:47:01.93 & 0.58 (0.02) & 0.099 & -8.726 (0.004) & 0.65 (0.01) & 0.11 & -8.432 (0.003) & -3.72 (0.06) \\
UCHII-OH4-1b & 17:20:53.371 &  -35:47:01.52 & 12.93 (0.02) & 2.2 & -10.699 (0.000) & 17.92 (0.04) & 3.0 & -10.304 (0.000) & -5.00 (0.01) \\
UCHII-OH6-1 & 17:20:53.397 &  -35:47:00.79 & 0.34 (0.00) & 0.058 & -10.153 (0.002) & 0.34 (0.00) & 0.058 & -10.491 (0.003) & +4.28 (0.05) \\
CM2-OH1-1 & 17:20:53.376 &  -35:46:55.64 & 2.98 (0.02) & 0.51 & -7.692 (0.001) & 2.67 (0.03) & 0.45 & -7.763 (0.003) & +0.90 (0.04) \\
CM2-OH1-2 & 17:20:53.391 &  -35:46:55.60 & 8.84 (0.07) & 1.50 & -7.723 (0.001) & 6.27 (0.11) & 1.1 & -7.799 (0.004) & +0.96 (0.05)
\enddata
\tablenotetext{a}{This is the mean of the intensity-weighted centroids of the associated RCP and LCP spots.  In all cases, they are within 1/3 of the synthesized beam minor axis of each other.}
\tablenotetext{b}{Values in parentheses show the fitted uncertainties.}
\tablenotetext{c}{Lower limits to the brightness temperature $T_B$ using the synthesized beam; the physical sizes of the maser spots are likely to be smaller than the beam ($\sim 520$~au).}
\end{deluxetable*}

The identification of Zeeman pairs can be challenging with an angular resolution of $0\farcs79\times 0\farcs25$ ($\sim 520$~au), i.e. significantly larger than the likely physical size of a single maser ``spot''.  We have nevertheless used the following procedure to kinematically and spatially separate (within the limitations of the data) the individual masing regions. For both transitions, we first identified maser ``spots'' of contiguous velocity components with fitted positions that were within 1/3 of the minor axis of the synthesized beam, independently for RCP and LCP. These independent RCP and LCP spot positions were then compared with each other and deemed to be a Zeeman pair if their intensity weighted centroid positions were within 1/3 of the minor axis of the synthesized beam.  Then the RCP and LCP profiles for each Zeeman pair were independently fitted using Gaussian profiles. In a few cases, spatially coincident Zeeman pairs consisted of more than one velocity component and these were further separated into independent ``spots'', though the spectral Gaussian fits per polarization were derived simultaneously. The derived parameters, including the fitted flux density and line center velocity ($V_{center}$) for the Zeeman pairs for both transitions are given in Table~\ref{blos}. The spots have been named according to their association, and then an additional number in increasing velocity order; spots with multiple velocity components (at the current resolution) are further appended with a letter of the alphabet in increasing velocity order. The line-of-sight magnetic field strength reported in Table~\ref{blos} was derived using $B_{los}=$($V_{center}(RCP) - V_{center}(LCP)$)$/Z$ where $Z = 0.056$~\kms\/mG$^{-1}$ for the 6.035~GHz transition and $0.079$~\kms\/mG$^{-1}$ for the 6.030~GHz transition \citep{Yen69}.

\begin{figure}[ht!]   
\includegraphics[width=0.99\linewidth]{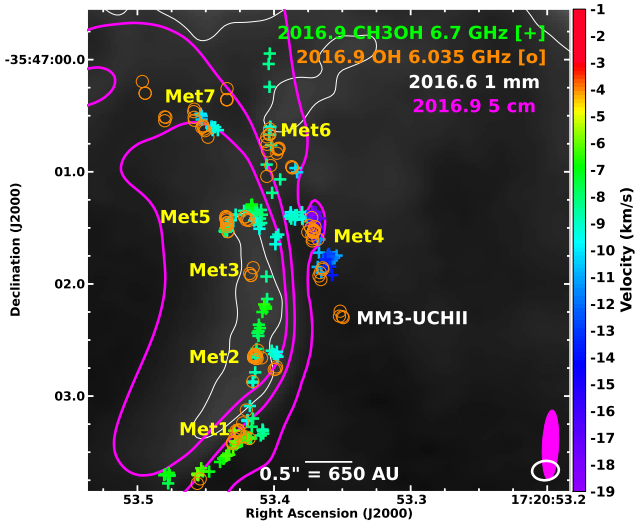}
\caption{Similar to Fig.~\ref{methanolzoom}d except the RCP OH 6.035~GHz maser spots (o symbols) are overplotted in orange for comparison with the positions of the methanol masers ($+$ symbols).
\label{methoh}}
\end{figure}

\citet{Caswell11} also report $B_{los}$ for these excited OH transitions from observations using three antennas of the LBA. Although these data have superior angular resolution ($0\farcs05\times 0\farcs02$), the absolute position accuracy was estimated to be as poor as $\pm0\farcs2$. In Figure~\ref{methoh} we have over-plotted the \citet{Caswell11} Zeeman pairs using $\diamond$ symbols, after applying a -0\farcs1 shift in right ascension in order to align them with the (more accurate) VLA Zeeman pairs from Table~\ref{blos}. From this figure, it is clear that toward the UCHII region many \citet{Caswell11} Zeeman  pairs are still present for both transitions, and with similar velocities.  Comparison of the derived $B_{los}$ between the present work and  \citet{Caswell11} also yields excellent overall agreement. For example, for spots that are in common between the two epochs at 6.035 GHz in the associations UCHII-OH2, UCHII-OH3, UCHII-OH4, UCHII-OH5, differences in $B_{los}$ are less than $10\%$. Only toward UCHII-OH6 are there notable differences in $B_{los}$ but this region also showed the largest variation in $B_{los}$ in the \citet{Caswell11} data. For the less prevalent 6.030 GHz transition, the agreement in the derived $B_{los}$ (for matching spots) with \citet{Caswell11} is also within $10\%$. Regarding our VLA  data, while most 6.030~GHz spots have derived $B_{los}$ that agree with the 6.035~GHz measurements, at a few locations the magnitude of the 6.030~GHz-derived field is notably larger (UCHII-OH2-1b, UCHII-OH2-2a, and especially CM2-OH1). This discrepancy may be a result of spectral/spatial blending for the stronger and more confused 6.035~GHz spots.

Additionally, we confirm the finding of \citet{Caswell11} of a reversal in the sign of $B_{los}$ between spots located toward the Southern portions (negative field: UCHII-OH1 to UCHII-OH5) and Northern portions (positive field: UCHII-OH6 and UCHII-OH7) of the \HII\/ region. Interestingly, the most northerly of the spots detected in UCHII-OH7 reveals a second reversal back to negative value \citep[this spot was not detected by][]{Caswell11}. Toward CM2-OH1, the $B_{los}$ is also positive, but somewhat weaker than for the majority of spots toward the \HII\/ region (as discussed previously, this is the first time excited OH masers have been resolved toward the CM2 region of the protocluster).

\subsection{Excited-state OH 4.660 GHz transition}

We did not detect the 4.660~GHz OH transition to a 
$3\sigma$ limit of 10~\mjb.  This transition is rarely 
seen in surveys \citep[see][for an earlier single-dish 
upper limit of 0.17~Jy]{Cohen95}. However, it was detected
at a peak flux density of 2.2~Jy in the field of \ngci\/ 
by the HartRAO single dish monitoring during the initial 
weeks of the flare, making it only the fifth object to 
have ever exhibited this transition \citep[see][and 
references therein]{MacLeod17}.  The emission velocity
of the transient 4.660~GHz maser was consistent with the flaring 
\methanol\/ and 6~GHz OH masers, suggesting that they were 
co-located.  But because no high-resolution observations 
exist, we lack definitive proof of this association.  


\section{Discussion}\label{disc}

\subsection{Implications from maser pumping schemes}

The rapid onset of methanol masers on and around MM1 raises
the question as to what conditions changed to support the maser inversion.
The 6.7~GHz methanol transition is radiatively pumped by mid-infrared photons { \citep{Sobolev97}}. The inversion requires \tdust\/ above ~120~K and can occur across a wide range of gas volume densities (up to about $n=10^{ 9}$~\ccc) and kinetic temperatures (at least 25 to 250~K), as long as the \methanol\/ abundance is greater than 10$^{-7}$ \citep{Cragg05}.  Prior to the millimeter continuum outburst, when the dust and gas temperatures were presumably in better equilibrium, the gas temperature provides an estimate for the dust temperature. In the line survey of \ngci\/ by \citet{Zernickel12}, the 
{ complex organic molecules with the most compact emission}
exhibited model excitation temperatures of 100~K (\methanol) to 150~K (CH$_3$CN).  Furthermore, the abundance of \methanol\/ was modeled as $4.7 \times 10^{-6}$. Following the millimeter continuum outburst, the dust temperature of MM1 inferred directly from the 1.3~mm continuum brightness temperature exceeds 250~K toward the central components, and was above 150~K over a several square arcsecond region \citep{Brogan16}.   Thus, in \ngci-MM1, the rapid heating of the dust grains by $\gtrapprox 100$~K by the recent accretion outburst can plausibly explain the appearance of 6.7~GHz masers in the gas in the vicinity of the powering source of the outburst, MM1B.  Furthermore, the lack of masers toward the central dust peaks (MM1A, 1B and 1D) suggests that the density is too high toward these objects to support the inversion, because the dust temperature is surely high enough.

\begin{figure}[ht!]   
\includegraphics[width=0.99\linewidth]{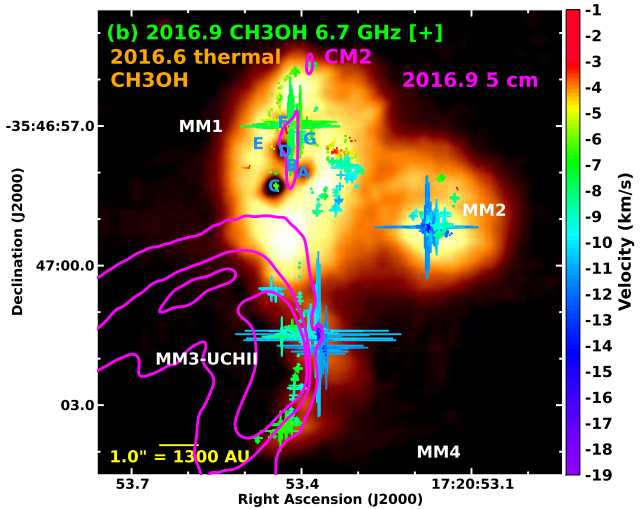}
\caption{The colorscale shows the peak intensity of the thermal methanol transition 11(2)-10(3) with an \Elower177~K. The same magenta contours, maser spot locations, and velocity color mapping from Fig.~\ref{methanolimage}b are also shown.  Blue letters denote the 1.3~mm continuum peaks in MM1 defined by \citet{Brogan16}.
\label{thermal}}
\end{figure} 

When compared to the ALMA 1~mm dust distribution (Fig.~\ref{methanolzoom}a), the appearance of 6.7~GHz masers to the north, west, and southwest of MM1 may seem perplexing, as the dust emission is relatively weak at all of these locations.
In fact, over 80\% of the masers reside outside of the 40\% level of the continuum, and over 50\% reside outside of the 5\% level of the continuum.
However, strong thermal molecular line emission is significantly more widespread than the bright dust emission. In Figure~\ref{thermal} we show the peak line intensity for a representative line from the ALMA 1~mm data (from whence the dust image also came). The transition shown is CH$_3$OH 11(2)-10(3) 
with an \Elower177~K (of order the expected gas kinetic temperature). From comparison of the distribution of 6.7~GHz masers versus the distribution of thermal CH$_3$OH (Fig.~\ref{thermal}), it is clear that the masers do lie in regions of high molecular column density.  The critical density ($n_{crit}$) of the CH$_3$OH 11(2)-10(3) transition can be computed from the ratio of the Einstein A coefficient ($3.46 \times 10^5$~s$^{-1}$) and the collisional cross section ($6\times10^{-13}$~cm$^{2}$) where the values are taken from \citet{Rabli10} via the LAMDA database \citep{Schoier10},  yielding $n_{crit} = 6\times10^7$~cm$^{-3}$.  { While the density at which strong emission can begin to arise from millimeter transitions \citep[the effective excitation density,][]{Shirley15} will be somewhat below $n_{crit}$, both values are} within the required range of $n$ for pumping the maser { \citep{Cragg05}}.  There is a notable lack of maser emission in the bright thermal gas between MM1 and MM3, suggesting that the gas density is too high here or the dust temperature is too low.

\begin{figure*}[ht!]   
\includegraphics[width=0.49\linewidth]{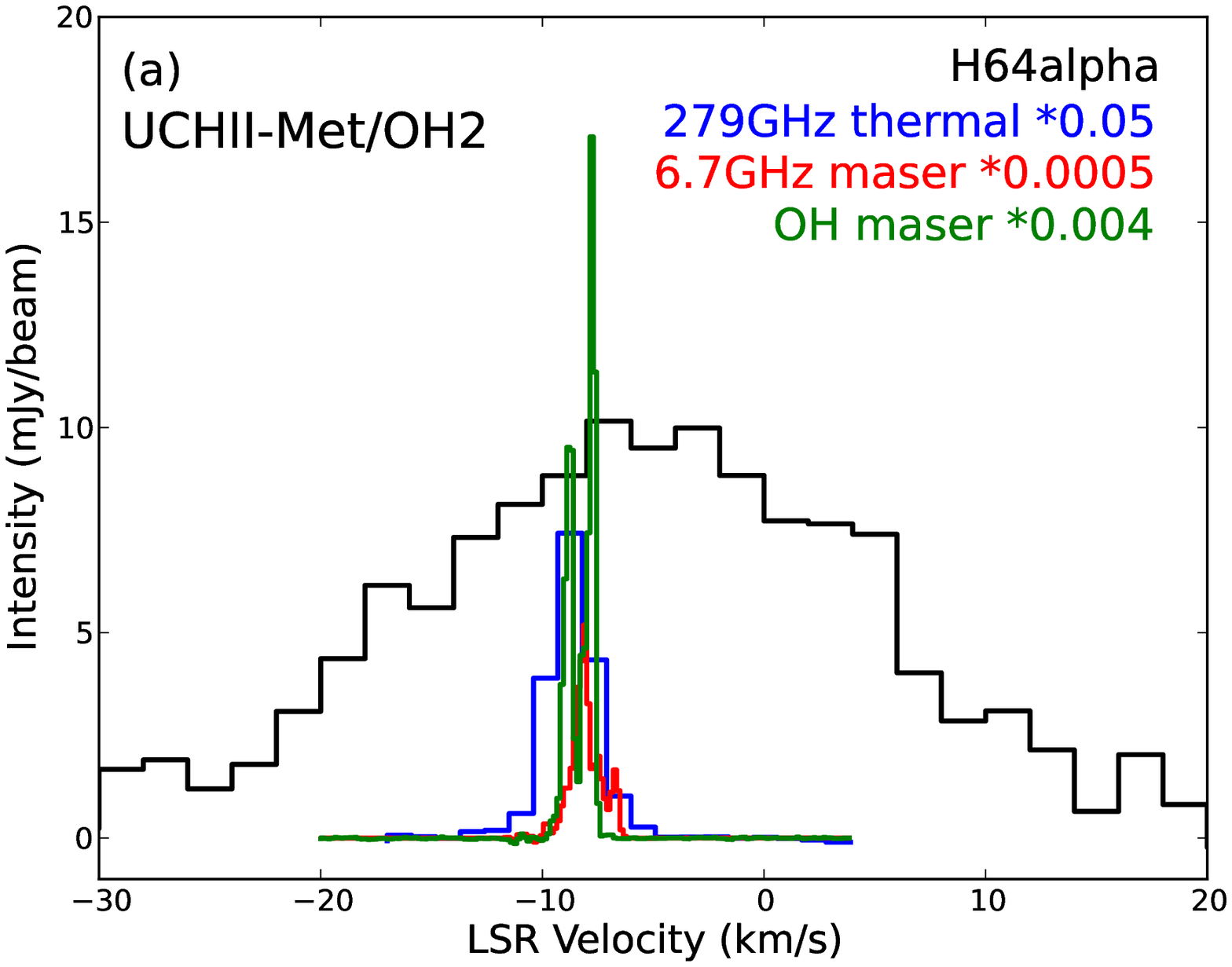}
\includegraphics[width=0.49\linewidth]{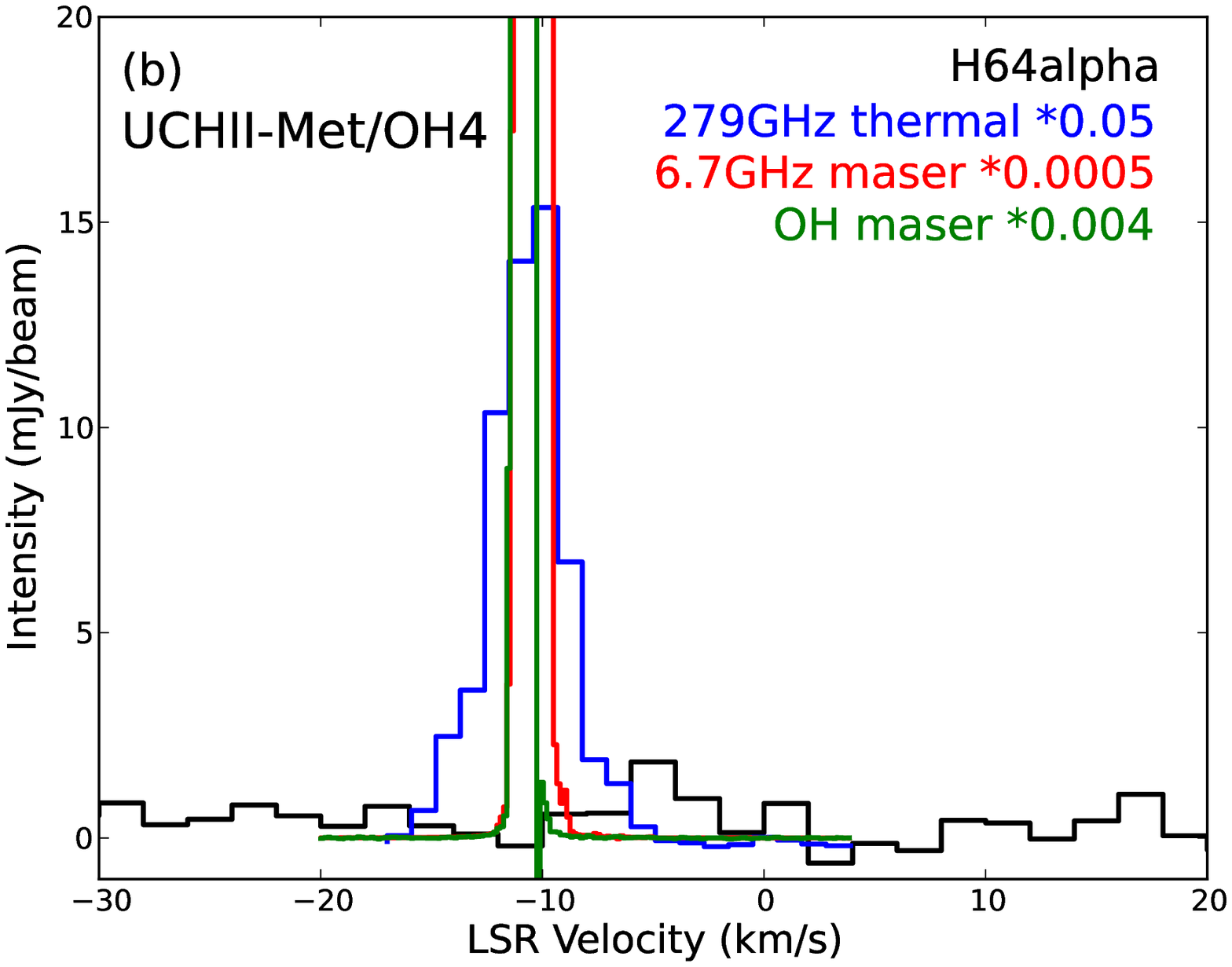}
\includegraphics[width=0.49\linewidth]{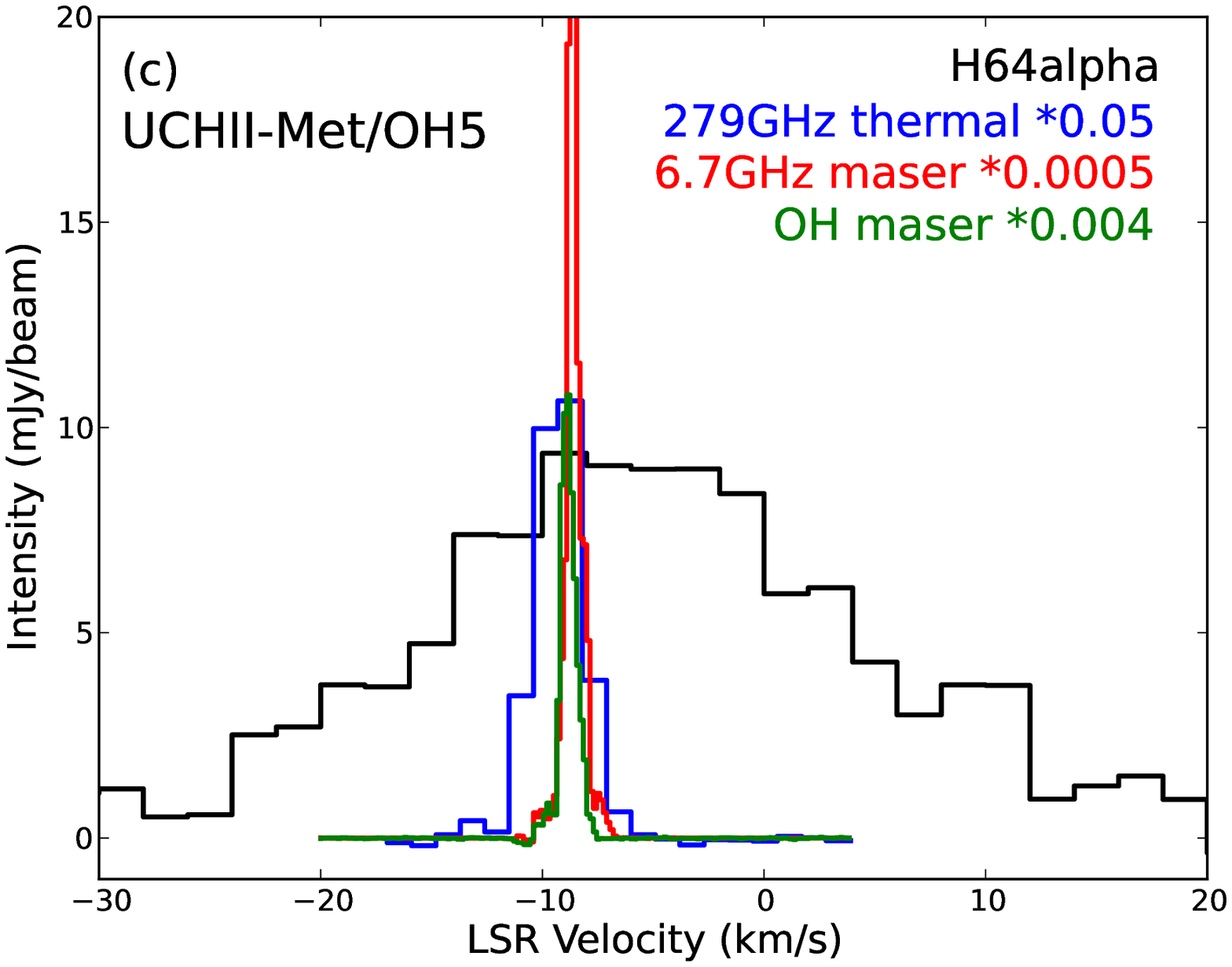}
\includegraphics[width=0.49\linewidth]{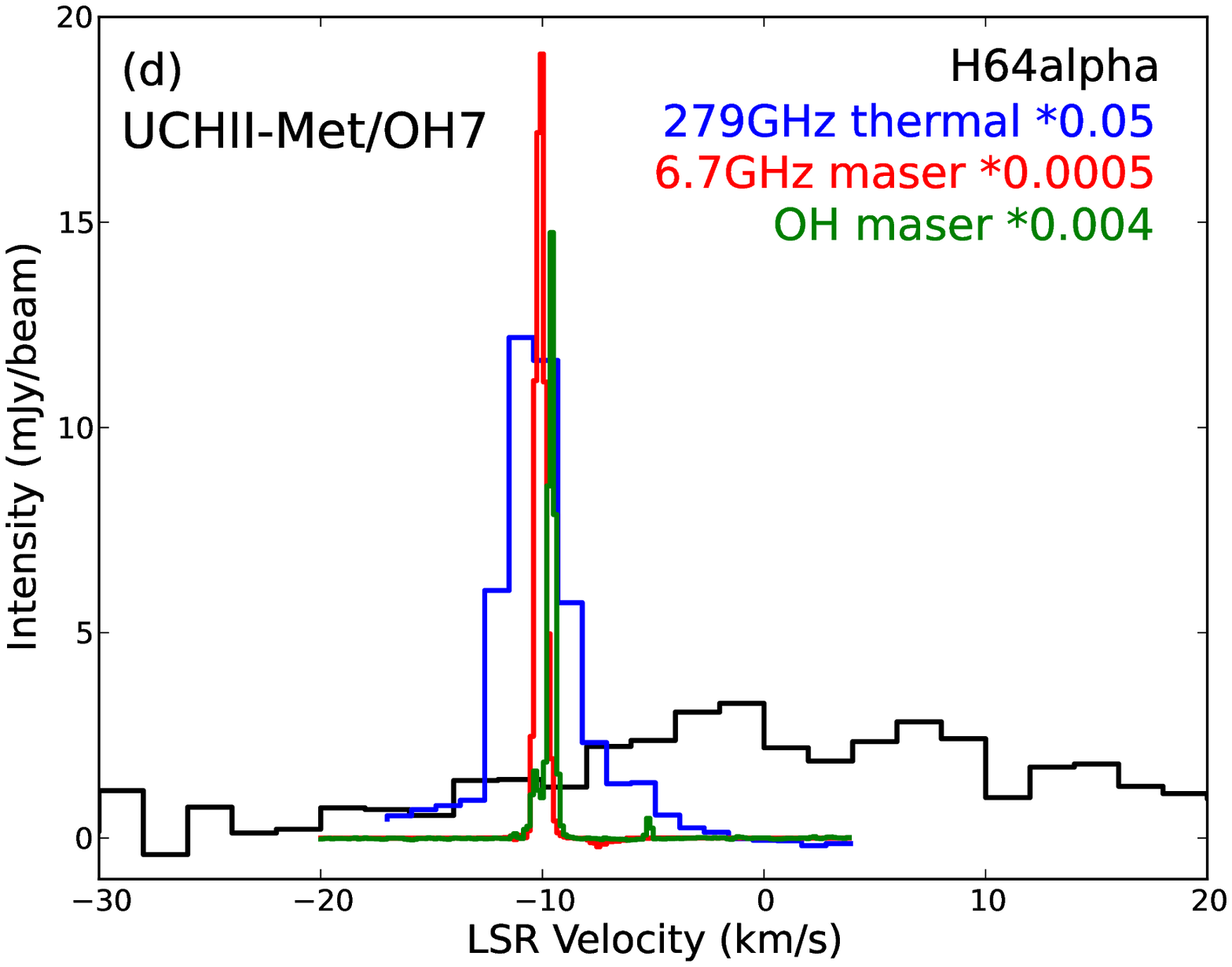}
\caption{Spectral profiles toward several of the maser associations in the MM3-UCHII region are shown.  The lines of methanol 6.7 and 279.352~GHz and of RCP OH 6.035 GHz are scaled down in intensity by the factor listed in the legend for ease of comparison to the weak recombination line.
\label{profiles}}
\end{figure*}

As demonstrated in Figure~\ref{methoh}, the 6.7~GHz and excited OH masers are coincident to within 
$<0\farcs1$ at a number of locations toward the MM3-UCHII region \citep[see also][]{Caswell97}, though it is notable that higher angular resolution studies generally find they are not exactly coincident at milliarcsecond scales, e.g. W3(OH) \citep{Etoka05}. 
Figure~\ref{profiles} shows example spectra comparing the velocities of the ionized gas (as traced by the H64$\alpha$ RRL), thermal gas (as traced by the CH$_3$OH 11(2,10)-10(3,7) transition), and 6.7 GHz and RCP for the 6.035~GHz maser gas toward several of the maser associations coincident with the MM3-UCHII region. Toward all of the maser associations (including those not shown), the velocities of both species of maser emission and of thermal \methanol\/ emission are in excellent agreement, suggesting that these gas components are also at similar locations along the line-of-sight. Toward the maser associations UCHII-Met/OH2, UCHII-Met/OH5, and UCHII-Met/OH7, the neutral gas is notably blueshifted with respect to the ionized gas. 
Assuming a typical expanding UCHII region scenario, this suggests that the neutral gas emission originates on the front side of MM3-UCHII. This effect is most dramatic toward the northern UCHII-Met/OH7 association where the neutral gas is blueshifted by $\gtrsim 10$ \kms\/ compared to the ionized gas, such that there is little overlap. It is notable that it is the ionized gas velocity that has changed compared to the other regions, rather than the maser velocities. The significant reddening of the ionized gas velocity toward the northern part of the MM3-UCHII region was also observed by \citet{dePree95} from VLA observations of the H76$\alpha$ line. The brightest 6.7~GHz and OH 6.035~GHz masers toward the UCHII region are found along its extreme western boundary (UCHII-Met/OH4), coincident with a $\sim 0\farcs25$ diameter blister-like bulge in the 5~cm continuum; unfortunately the H64$\alpha$ line was not detected at this location, so the relationship between the masers and ionized gas cannot be determined. The RRL was not detected toward the CM2 region { (nor toward MM1)}, though as for the UCHII region, the positions and velocities of the { CM2} 6.7~GHz and OH 6.035~GHz masers are similar, and their peak velocities match the thermal methanol that is present there.

Excited state OH masers are efficiently pumped for a lower range of kinetic (gas) temperatures than the 6.7~GHz \methanol\/ transition: 25 - 70~K \citep[see for example][]{Cragg02}. Therefore, the excellent kinematic agreement between the two maser species suggests that the molecular gas toward the MM3-UCHII region is cooler than toward MM1 (where excited OH masers are not detected). The maser models of \citet{Cragg02,Cragg05} also suggest that the excited 6.030 and 6.035~GHz OH transitions are only effectively pumped toward the upper end of the 
density range predicted for 6.7~GHz \methanol\/ masers ($10^{6.5}$ to $10^{8.3}$~\ccc\/ compared to $10^{4}$ to $10^{8.3}$~\ccc). Interestingly, the \citet{Cragg02} models also suggest that the ratio of 6.035/6.030~GHz OH maser brightness temperature is minimized, and indeed becomes nearly equal, for the highest densities, which may explain the anomalous ratio observed toward CM2 if the gas density is higher there. Future multi-transition analysis of the thermal molecular gas throughout the protocluster will help to verify that the observed physical conditions meet the predictions of maser pump models. 

{ The range of the magnetic field strengths we measure in excited OH (0.5-5~mG) is very similar to that found in other high-mass star
forming regions observed interferometrically in these 
transitions \citep[e.g., W51~Main, W3~(OH), and ON1,][]{Etoka12,Fish07,Green07}. 
These values are about five times higher than those measured in
single-dish CN observations of massive star forming regions
\citep{Falgarone08}, however those observations sample gas at larger scales and hence lower average density (1-20~$\times 10^5$~\ccc). Because magnetic field strength is expected to scale with the square root of density \citep{Crutcher99}, the excited OH masers apparently arise from  gas at densities that 
are $\sim$25 times higher, i.e. 0.2-5~$\times 10^7$~\ccc.  
Radiative transfer modeling of these transitions predict that OH densities of $\sim$10~\ccc\/ are required to pump these masers, which are consistent with H$_2$ densities of $5 \times 10^7$~\ccc\/ using the nominal OH abundance of $2 \times 10^{-7}$ \citep{Etoka12}.  Thus, the density implied by the measured excited OH field strengths is in good agreement with the value required for the thermal methanol emission and the pumping models for excited OH maser emission.}

\subsection{Spatial and kinematic structures}

The sudden appearance of strong 6.7~GHz methanol masers on and around MM1 demonstrates an important new characteristic of the Class II maser phenomenon, which has presented an ongoing enigma to observers. 
While Class I methanol masers \citep{Leurini16} trace ambient-velocity shocked gas 
at the interfaces of outflow structures and can be located far from the driving source 
\citep{Cyganowski09,Rodriguez17}, Class II methanol masers are always found in close proximity to other tracers of massive protostars such as submillimeter continuum \citep{Urquhart13}, 
mid-infrared continuum \citep{Bartkiewicz14}, or hot core molecular gas \citep{Chibueze17}.  
An early imaging survey with the ATCA at $\sim1''$ resolution found linear distributions of 
fitted spots across (up to) a few synthesized beams, which were interpreted as a indicator of 
either edge-on protoplanetary disks or collimated jets \citep{Norris93}.  Interestingly, two 
of the 10 so-called ``linear'' sources were the two concentrations of emission in NGC6334F 
(the previously-known masers associated with MM2 and MM3). Although only half of the cases exhibited a corresponding velocity gradient, the edge-on disk interpretation became preferred in the literature \citep{Norris98}. 

However, the disk hypothesis was subsequently tested
by searching for outflow lobes perpendicular to the linear structures by imaging the shock-tracing near-infrared H$_2$ line in 28 such maser sources \citep{DeBuizer03}.  Only two sources showed outflowing gas in the perpendicular direction, while in contrast 80\% of the cases (with radiatively excited H$_2$) showed emission parallel to the maser axis, instead suggesting an association with the 
outflow.  Further evidence for an association between 6.7~GHz masers and outflows has come from other studies of individual sources \citep[e.g. NGC7538~IRS1,][]{DeBuizer05}. 
Nevertheless, the disk interpretation 
continues to be invoked to explain recent observations of individual massive protostars, such 
as Cepheus~A~HW2 \citep{Sanna17} and G353.273+0.641 \citep{Motogi17}; however in these cases, 
the proposed disks have an inclined or face-on geometry.  In one case, proper motion studies suggest masers reside in both the disk and outflow \citep[IRAS~20126+4104,][]{Moscadelli11}.
Recently, a VLBI imaging survey of 31 
methanol masers identified a new class of structure: a ring-like distribution of spots in 29\% of the sample \citep{Bartkiewicz09}, with the most striking example being G23.657-0.127 
\citep{Bartkiewicz05}.   With typical radii of $\lesssim0\farcs1$, these rings are smaller 
than the scales probed by the VLA or ATCA. A follow-up study in the near-IR and mid-IR again 
found no evidence to support the idea that these rings trace protostellar disks 
\citep{DeBuizer12}. Proper motion studies  show that the ring of masers in G23.657-0.127 is expanding \citep{Szymczak17} and while the interpretation is still unclear, it may be the result of a disk wind  \citep{Bartkiewicz17}.

{ Before interpreting the location of the flaring masers, we summarize the current prior knowledge of star formation in MM1.  The original ALMA observations resolved MM1 into multiple dust continuum components, which we modeled as seven two-dimensional Gaussian sources at 1.3~mm \citep[A-F,][]{Brogan16}.  Since the molecular hot core emission encompasses all these components, each one should be considered as a candidate protostar.  However, the water maser emission from MM1 (VLA epoch 2011.7) was associated with only the two brightest components, B and D.  The small upper limit to the size of B at 7~mm ($<$0\farcs18 $\sim230$~au) led us to model its spectral energy distribution (SED) as a dense hypercompact HII (HCHII) region, while the SED and larger size of D were more consistent with emission from a jet.  A single component fit to the elongated 5~cm emission feature (epoch 2011.4) peaks near D, but extends down to B and up to an area between F and G.  We interpreted this emission as a jet, but its origin and whether it is symmetric with respect to its driving source (D) or more asymmetric (B) were unclear.  Besides B and D, the only other source detected at 7~mm is F, whose SED is consistent with dust emission alone, suggesting an earlier stage protostar. The other 5~cm source, CM2, was enigmatic as its emission is non-thermal and it exhibits the brightest water maser emission yet no compact dust emission.}

The distribution of the new flaring masers in MM1 exhibits a variety of shapes.  There are a few linear arrangements of spots, perhaps the most striking of which is the ``V''-shaped structure of masers between MM1F and 1G (Fig.~\ref{methanolzoom}b). Because this structure lies at the northern end of the 5~cm continuum emission, { and along an axis parallel to this emission that intersects MM1B}, it may indicate an association between the masers and a jet originating from { the HCHII region} MM1B.  { This axis also passes close to the non-thermal radio source CM2, suggesting that it traces a shock against the ambient medium, as proposed for the non-thermal component in the HH~80-81 jet \citep{RodriguezKamenetzky}.
How this jet relates to the large scale NE/SW bipolar outflow remains uncertain from these data, but we note a similarity to the G5.89-0.39 protocluster, which exhibits a young compact outflow  and an older larger scale outflow at very different position angles \citep{Hunter08,Puga06}.}
While it is { also} unclear if the maser spots near MM1F are associated with an { early stage} protostar at that location, the thread of spots that continues northward (CM2-Met2) curving toward the non-thermal continuum source CM2 (Fig.~\ref{methanolzoom}a) supports an association with the jet.  The spots north of MM1G partially form a similar structure west of the jet, which suggests that both sets of masers follow the walls of an outflow cavity.  
Proper motion studies with VLBI would be helpful to test this hypothesis { (Chibueze et al., in preparation).}

Shorter linear collections of maser spots also appear in MM1-MetC and MM1-Met3.  The dust emission from MM1C indicates there may be a protostar associated with MM1-MetC, however the velocity range of these spots is rather small (2.7~\kms) and lacks a systematic gradient.  MM1-Met3 does show an east/west velocity gradient, but the velocity range is even smaller (1.3~\kms) and there is no evidence for a protostar as a point source in the dust emission. Linear features with such narrow velocity ranges are similar to those seen in LBA images of five other HMYSOs by \citet{Dodson04}, who conclude that they trace planar shocks propagating perpendicular to the line of sight rather than disks.  Similar to MM1-Met3, the copious maser spots of MM1-Met2 have no direct association with a protostar, since there are no millimeter point sources in that area to very sensitive levels \citep{Brogan16}. Instead, these maser spots look more like they arise from a collection of filaments, similar to those seen in W3OH where masers trace filaments extending up to 3100~au \citep{HarveySmith06}.  It would appear that the presence of maser filaments can occur over a broad range in evolutionary state of massive protostars since the masers in MM3-UCHII (Met1 through Met7) appear to delineate a filament nearly 5000~au in length (Fig.~\ref{methoh}).

In the other HMYSO hosting methanol masers in this protocluster, MM2, we see a similar dichotomy as in MM1, with the spots of MM2-Met1 appearing close to the protostar(s) traced by the continuum sources MM2A and 2B, while those of MM2-Met2 appear further afield ($>1000$~au). So the primary conclusion we draw from this outburst is that although some 6.7~GHz masers do reside close to massive protostars (within 500~au), significant amounts of maser emission can occur in the gas further away as long as the radiative pumping from nearby protostars is sufficient in those locations.  The presence of outflow cavities driven by a central protostar likely creates propagation paths for infrared photons that increase the range of the required radiative pumping to these larger distances.  Indeed, the brightest masers can lie along these paths, as evidenced by MM1-Met1, which lies $\approx$1000~au from the outburst source MM1B.  This result is strikingly similar to the other recent flaring maser source S255~NIRS3, in which the brightest maser emission also lies 500-1000~au from the flaring protostar \citep{Moscadelli17}.


\subsection{Origin of the 1999 methanol maser flare}

\citet{Goedhart04} presented single-dish monitoring data of the 6.7~GHz line from 28 February 1999 to 27 March 2003 and noted a large flare in the velocity component at $-5.88$~\kms. Re-analysis
of these data \citep{MacLeod17} indicates that the flare peaked on 19 November 1999 (epoch 1999.88) at a velocity of $-5.99$~\kms\/ with a  peak flux density 230 times higher than the pre-burst value of $1.6\pm0.4$~Jy.  While there are no published interferometric maps from that epoch, we can examine the location of the emission at this velocity in our recent epoch 2016.9 VLA data.  In the $-5.90$~\kms\/ channel, we find a total flux density of 6.41~Jy, only 10\% of which originates from any of the UCHII associations (only UCHII-Met1).  The majority (5.41~Jy) originates from MM1, and the rest (0.34~Jy) from CM2-Met2.   Since the rest of the emission (0.70~Jy, from UCHII-Met1 and MM2-Met2) is consistent with the quiescent single-dish value found after the 1999 flare ($0.55\pm0.20$~Jy), it is quite plausible that this flare could have also originated entirely from MM1 and CM2.  This result supports the hypothesis of \citet{MacLeod17} that the 1999 and 2015 events arise from the same physical location and are due to a common repeating mechanism such as the decaying orbit of a binary system embedded in dense gas \citep{Stahler10}.  Continued long-term monitoring of the maser emission is essential to test this scenario.  It is particularly important to measure how long the current maser flare lasts compared to the more limited 1999 event.  If this flare persists for many years, it increases the likelihood that a significant population of the masers found in surveys arise from protostars that have recently undergone a large accretion outburst.

\section{Conclusions}

The recent extraordinary outburst in the massive protostellar cluster NGC6334I provides an unprecedented window into the central environment of a massive star-forming region.  We have obtained high resolution ($\sim$500~au) VLA images of multiple maser transitions which reveal the first appearance of Class II methanol maser emission toward the MM1 protostellar system in over 30 years of past observations.  The 6.7~GHz masers are distributed toward particular parts of MM1 (including MM1C, F and G), on its northern and western peripheries, and toward the non-thermal radio source CM2.  The masers are strikingly absent from the brightest millimeter components (MM1A, B and D). We find that current models of maser pumping can explain the general location of the masers in the context of the recent luminous outburst from MM1B.  The spatio-kinematic structures traced by the 6.7~GHz masers are varied but support the idea that while some masers do reside close to massive protostars (within 500~au) traced by compact dust emission, significant maser action can also occur in more extended areas associated with strong thermal gas emission.   Also, the presence of jets and outflow cavities driven by a central protostar increases the range of infrared pumping photons, allowing strong masers to appear further away ($>1000$~au).

Our simultaneous observations of two of the 6~GHz excited state OH transitions reveal no emission toward MM1, but we do detect a strong new maser at $-7.7$~\kms\/ arising from the non-thermal radio source CM2.  Not previously detected by the LBA, it was apparently excited by the recent millimeter outburst.  By analyzing the RCP and LCP data cubes, we identify several Zeeman pairs to measure the line of sight magnetic field in the UCHII region and CM2. We confirm the field strengths and the reversal in field direction across the UCHII region identified by \citet{Caswell11}.   The magnetic field toward CM2 is significantly weaker than toward the majority of positions across the UCHII region, and the flux ratio between the 6.035 and 6.030~GHz lines is anomalous, with the lines being comparable in strength toward CM2.  OH maser pumping models suggest that the gas density is significantly higher toward CM2, consistent with compression of gas by the propagation of the jet from MM1B.  

Future high resolution observations of the \methanol\/ and OH masers will be essential to obtain proper motions of the new features in order to determine which ones may be tracing structures associated with an outflow versus those that may be bound to a protostar. Comparison with thermal gas outflow tracers imaged by ALMA will then be possible.  Continued single dish monitoring is also critical to measure the lifetime of the current maser flare and to search for future events in order to test the hypothesis that a periodic phenomenon such as an eccentric protostellar binary orbit may be responsible for repeating accretion outbursts.

\acknowledgments

{ We thank the anonymous referee for a thorough review which has improved the manuscript.} The National Radio Astronomy Observatory is a facility of the National Science Foundation operated under agreement by the Associated Universities, Inc. 
This paper makes use of the following ALMA data: ADS/JAO.ALMA\#2015.A.00022.T. ALMA is a partnership of ESO (representing its member states), NSF (USA) and NINS (Japan), together with NRC (Canada) and NSC and ASIAA (Taiwan) and KASI (Republic of Korea), in cooperation with the Republic of Chile. The Joint ALMA Observatory is operated by ESO, AUI/NRAO and NAOJ.   This research made use of NASA's Astrophysics Data System Bibliographic Services, the SIMBAD database operated at CDS, Strasbourg, France, { Astropy, a community-developed core Python package for Astronomy \citep{astropy}, and APLpy, an open-source plotting package for Python hosted at http://aplpy.github.com. T. Hirota is supported by the MEXT/JSPS KAKENHI grant No. 17K05398.}  C.J.~Cyganowski acknowledges support from the STFC (grant number ST/M001296/1).

Facilities: \facility{VLA}, \facility{ALMA}.

\clearpage


\begin{thebibliography}{}

\bibitem[Abraham et al.(1981)]{Abraham81} Abraham, Z., Cohen, N.~L., Opher, R., Raffaelli, J.~C., \& Zisk, S.~H.\ 1981, \aap, 100, L10 

\bibitem[Astropy Collaboration et al.(2013)]{astropy} 
{ Astropy Collaboration, Robitaille, T.~P., Tollerud, E.~J., et al.\ 2013, \aap, 558, A33 }

\bibitem[Avison et al.(2016)]{Avison16} Avison, A., Quinn, L.~J., Fuller, G.~A., et al.\ 2016, \mnras, 461, 136 




\bibitem[Bartkiewicz et al.(2017)]{Bartkiewicz17} Bartkiewicz, A., et al. 2017, IAU Symp. 336

\bibitem[Bartkiewicz et al.(2014)]{Bartkiewicz14} Bartkiewicz, A., Szymczak, M., \& van Langevelde, H.~J.\ 2014, \aap, 564, A110 

\bibitem[Bartkiewicz et al.(2009)]{Bartkiewicz09} Bartkiewicz, A., Szymczak, M., van Langevelde, H.~J., Richards, A.~M.~S., \& Pihlstr{\"o}m, Y.~M.\ 2009, \aap, 502, 155 

\bibitem[Bartkiewicz et al.(2005)]{Bartkiewicz05} Bartkiewicz, A., Szymczak, M., \& van Langevelde, H.~J.\ 2005, \aap, 442, L61 

\bibitem[Beuther et al.(2008)]{Beuther08}
{ Beuther, H., Walsh, A.~J., Thorwirth, S., et al.\ 2008, \aap, 481, 169 }

\bibitem[Beuther et al.(2007)]{Beuther07} Beuther, H., Walsh, A.~J., Thorwirth, S., et al.\ 2007, \aap, 466, 989 





\bibitem[Brogan et al.(2016)]{Brogan16} Brogan, C.~L., Hunter, T.~R., Cyganowski, C.~J., et al.\ 2016, \apj, 832, 187


\bibitem[Caratti o Garatti et al.(2017)]{Caratti16} Caratti o Garatti, A., Stecklum, B., Garcia Lopez, R., et al.\ 2017, Nature Physics, 13, 276 



\bibitem[Caswell \& Vaile(1995)]{Caswell95} Caswell, J.~L., \& Vaile, R.~A.\ 1995, \mnras, 273, 328 

\bibitem[Caswell(1997)]{Caswell97} Caswell, J.~L.\ 1997, \mnras, 289, 203 

\bibitem[Caswell et al.(2011)]{Caswell11} Caswell, J.~L., Kramer, B.~H., \& Reynolds, J.~E.\ 2011, \mnras, 414, 1914 

\bibitem[Caswell(2003)]{Caswell03} Caswell, J.~L.\ 2003, \mnras, 341, 551 




\bibitem[Chibueze et al.(2017)]{Chibueze17} Chibueze, J.~O., Csengeri, T., Tatematsu, K., et al.\ 2017, \apj, 836, 59 

\bibitem[Chibueze et al.(2014)]{Chibueze14}
Chibueze, J.~O., Omodaka, T., Handa, T., et al.\ 2014, \apj, 784, 114 


\bibitem[Cohen et al.(1995)]{Cohen95} Cohen, R.~J., Masheder, M.~R.~W., \& Caswell, J.~L.\ 1995, \mnras, 274, 808 

\bibitem[Contreras Pe{\~n}a et al.(2017)]{Contreras17} Contreras Pe{\~n}a, C., Lucas, P.~W., Minniti, D., et al.\ 2017, \mnras, 465, 3011 

\bibitem[Cragg et al.(2005)]{Cragg05} Cragg, D.~M., Sobolev, A.~M., \& Godfrey, P.~D.\ 2005, \mnras, 360, 533 

\bibitem[Cragg et al.(2002)]{Cragg02} Cragg, D.~M., Sobolev, A.~M., \& Godfrey, P.~D.\ 2002, \mnras, 331, 521 


\bibitem[Crutcher(1999)]{Crutcher99} 
{ Crutcher, R.~M.\ 1999, \apj, 520, 706 }

\bibitem[Cyganowski et al.(2009)]{Cyganowski09} Cyganowski, C.~J., Brogan, C.~L., Hunter, T.~R., \& Churchwell, E.\ 2009, \apj, 702, 1615 



\bibitem[De Buizer et al.(2012)]{DeBuizer12} De Buizer, J.~M., Bartkiewicz, A., \& Szymczak, M.\ 2012, \apj, 754, 149 

\bibitem[De Buizer \& Minier(2005)]{DeBuizer05} De Buizer, J.~M., \& Minier, V.\ 2005, \apjl, 628, L151 

\bibitem[De Buizer(2003)]{DeBuizer03} De Buizer, J.~M.\ 2003, \mnras, 341, 277 


\bibitem[De Buizer et al.(2002)]{DeBuizer02} De Buizer, J.~M., Radomski, J.~T., Pi{\~n}a, R.~K., \& Telesco, C.~M.\ 2002, \apj, 580, 305 



\bibitem[de Pree et al.(1995)]{dePree95} de Pree, C.~G., Rodriguez, L.~F., Dickel, H.~R., \& Goss, W.~M.\ 1995, \apj, 447, 220 



\bibitem[Dodson \& Moriarty(2012)]{Dodson12} Dodson, R., \& Moriarty, C.~D.\ 2012, \mnras, 421, 2395 

\bibitem[Dodson et al.(2004)]{Dodson04} Dodson, R., Ojha, R., \& Ellingsen, S.~P.\ 2004, \mnras, 351, 779 


\bibitem[Elitzur(1992)]{Elitzur92} Elitzur, M.\ 1992, \araa, 30, 75 
\bibitem[Ellerbroek et al.(2011)]{Ellerbroek11} 
{ Ellerbroek, L.~E., Kaper, L., Bik, A., et al.\ 2011, \apjl, 732, L9 }

\bibitem[Ellingsen(1996)]{Ellingsen96} Ellingsen, S.\ 1996, Ph.D.~Thesis,  University of Tasmania

\bibitem[Ellingsen et al.(1996)]{Ellingsen96etal} Ellingsen, S.~P., Norris, R.~P., Diamond, P.~J., et al.\ 1996, Proc. of Third Asia-Pacific Telescope Workshop, arXiv:astro-ph/9604024 


\bibitem[Etoka et al.(2012)]{Etoka12} 
{ Etoka, S., Gray, M.~D., \& Fuller, G.~A.\ 2012, \mnras, 423, 647 }

\bibitem[Etoka et al.(2005)]{Etoka05} Etoka, S., Cohen, R.~J., \& Gray, M.~D.\ 2005, \mnras, 360, 1162 

\bibitem[Evans et al.(2009)]{Evans09} Evans, N.~J., II, Dunham, M.~M., J{\o}rgensen, J.~K., et al.\ 2009, \apjs, 181, 321-350 

\bibitem[Falgarone et al.(2008)]{Falgarone08} 
{ Falgarone, E., Troland, T.~H., Crutcher, R.~M., \& Paubert, G.\ 2008, \aap, 487, 247 }


\bibitem[Fish \& Sjouwerman(2007)]{Fish07} Fish, V.~L., \& Sjouwerman, L.~O.\ 2007, \apj, 668, 331 



\bibitem[Fujisawa et al.(2015)]{Fujisawa15} Fujisawa, K., Yonekura, Y., Sugiyama, K., et al.\ 2015, The Astronomer's Telegram, 8286

\bibitem[Fujisawa et al.(2012)]{Fujisawa12} Fujisawa, K., Sugiyama, K., Aoki, N., et al.\ 2012, \pasj, 64, 17 


\bibitem[Gardner et al.(1970)]{Gardner70} Gardner, F.~F., Ribes, J.~C., \& Goss, W.~M.\ 1970, \aplett, 7, 51 

\bibitem[Goedhart et al.(2004)]{Goedhart04} Goedhart, S., Gaylard, M.~J., \& van der Walt, D.~J.\ 2004, \mnras, 355, 553 

\bibitem[Gray et al.(2016)]{Gray16} Gray, M.~D., Baudry, A., Richards, A.~M.~S., et al.\ 2016, \mnras, 456, 374 

\bibitem[Green et al.(2015)]{Green15} Green, J.~A., Caswell, J.~L., \& McClure-Griffiths, N.~M.\ 2015, \mnras, 451, 74 

\bibitem[Green et al.(2007)]{Green07} 
{ Green, J.~A., Richards, A.~M.~S., Vlemmings, W.~H.~T., Diamond, P., \& Cohen, R.~J.\ 2007, \mnras, 382, 770 }




\bibitem[Harvey-Smith \& Cohen(2006)]{HarveySmith06} Harvey-Smith, L., \& Cohen, R.~J.\ 2006, \mnras, 371, 1550 

\bibitem[Herbig(1977)]{Herbig77} Herbig, G.~H.\ 1977, \apj, 217, 693 

\bibitem[Herbig(1989)]{Herbig89} Herbig, G.~H.\ 1989, European Southern Observatory Conference and Workshop Proceedings, 33, 233 

\bibitem[Hirota et al.(2014)]{Hirota14} Hirota, T., Tsuboi, M., Kurono, Y., et al.\ 2014, \pasj, 66, 106 




\bibitem[Honma et al.(2004)]{Honma04} Honma, M., Yoon, K.~C., Bushimata, T., et al.\ 2004, \pasj, 56, L15 


\bibitem[Hunter et al.(2017)]{Hunter17} Hunter, T.~R., Brogan, C.~L., MacLeod, G., et al.\ 2017, \apjl, 837, L29 


\bibitem[Hunter et al.(2008)]{Hunter08} 
{ Hunter, T.~R., Brogan, C.~L., Indebetouw, R., \& Cyganowski, C.~J.\ 2008, \apj, 680, 1271 }

\bibitem[Hunter et~al.(2006)]{Hunter06}
{Hunter}, T.~R., {Brogan}, C.~L., {Megeath}, et al.,
2006, \apj, 649, 888




\bibitem[Kenyon et al.(1990)]{Kenyon90} Kenyon, S.~J., Hartmann, L.~W., Strom, K.~M., \& Strom, S.~E.\ 1990, \aj, 99, 869 

\bibitem[Knowles et al.(1976)]{Knowles76} Knowles, S.~H., Caswell, J.~L., \& Goss, W.~M.\ 1976, \mnras, 175, 537 

\bibitem[Knowles et al.(1973)]{Knowles73} Knowles, S.~H., Johnson, K.~J., Moran, J.~M., \& Ball, J.~A.\ 1973, \apjl, 180, L117 


%

\bibitem[Krishnan et al.(2013)]{Krishnan13} Krishnan, V., Ellingsen, S.~P., Voronkov, M.~A., \& Breen, S.~L.\ 2013, \mnras, 433, 3346 

\bibitem[Kumar et al.(2016)]{Kumar16} Kumar, M.~S.~N., Contreras Pe{\~n}a, C., Lucas, P.~W., \& Thompson, M.~A.\ 2016, \apj, 833, 24 

\bibitem[Lekht et al.(2017)]{Lekht17} 
{ Lekht, E.~E., Pashchenko, M.~I., Rudnitskij, G.~M., \& Tolmachev, A.~M.\ 2017, Astronomy Reports, submitted, arXiv:1709.08197 }

\bibitem[Leurini et al.(2016)]{Leurini16} Leurini, S., Menten, K.~M., \& Walmsley, C.~M.\ 2016, \aap, 592, A31 

\bibitem[Leurini et~al.(2006)]{Leurini06}
{ {Leurini}, S., {Schilke}, P., {Parise}, B., et al., 2006, \aap, 454, L83}

\bibitem[Liljestr{\"o}m \& Gwinn(2000)]{Liljestrom00} Liljestr{\"o}m, T., \& Gwinn, C.~R.\ 2000, \apj, 534, 781 





\bibitem[Lucas et al.(2017)]{Lucas17} Lucas, P.~W., Smith, L.~C., Contreras Pena, C., et al.\ 2017, \mnras, 472, 2990

\bibitem[Lucas et al.(2008)]{Lucas08} Lucas, P.~W., Hoare, M.~G., Longmore, A., et al.\ 2008, \mnras, 391, 136 

\bibitem[MacLeod et al.(2018)]{MacLeod17}
Macleod, G. et al., { 2018}, submitted to MNRAS 

\bibitem[Mairs et al.(2017)]{Mairs17} Mairs, S., Lane, J., Johnstone, D., et al.\ 2017, \apj, 843, 55 

\bibitem[McCutcheon et al.(2000)]{McCutcheon00} 
{ McCutcheon, W.~H., Sandell, G., Matthews, H.~E., et al.\ 2000, \mnras, 316, 152 }


\bibitem[McGuire et al.(2017)]{McGuire17}
McGuire, B. A., et al.\ 2017, \apjl, 851, L46 

\bibitem[Meyer et al.(2017)]{Meyer17} Meyer, D.~M.-A., Vorobyov, E.~I., Kuiper, R., \& Kley, W.\ 2017, \mnras, 464, L90 


\bibitem[Minier et al.(2002)]{Minier02} Minier, V., Booth, R.~S., \& Conway, J.~E.\ 2002, \aap, 383, 614 



\bibitem[Minniti et al.(2010)]{Minniti10} Minniti, D., Lucas, P.~W., Emerson, J.~P., et al.\ 2010, New Astronomy, 15, 433 

\bibitem[Mitchell et al.(1991)]{Mitchell91} 
{ Mitchell, G.~F., Maillard, J.-P., \& Hasegawa, T.~I.\ 1991, \apj, 371, 342 }

\bibitem[Moscadelli et al.(2017)]{Moscadelli17} Moscadelli, L., Sanna, A., Goddi, C., et al.\ 2017, \aap, 600, L8 

\bibitem[Moscadelli et al.(2011)]{Moscadelli11} Moscadelli, L., Cesaroni, R., Rioja, M.~J., Dodson, R., \& Reid, M.~J.\ 2011, \aap, 526, A66 

\bibitem[Motogi et al.(2017)]{Motogi17} Motogi, K., Hirota, T., Sorai, K., et al.\ 2017, \apj, 849, 23

\bibitem[Norris et al.(1988)]{Norris88} Norris, R.~P., Caswell, J.~L., Wellington, K.~J., McCutcheon, W.~H., \& Reynolds, J.~E.\ 1988, \nat, 335, 149 


\bibitem[Norris et al.(1993)]{Norris93} Norris, R.~P., Whiteoak, J.~B., Caswell, J.~L., Wieringa, M.~H., \& Gough, R.~G.\ 1993, \apj, 412, 222 

\bibitem[Norris et al.(1998)]{Norris98} Norris, R.~P., Byleveld, S.~E., Diamond, P.~J., et al.\ 1998, \apj, 508, 275 






\bibitem[Offner \& McKee(2011)]{Offner11} Offner, S.~S.~R., \& McKee, C.~F.\ 2011, \apj, 736, 53 


\bibitem[Omodaka et al.(1999)]{Omodaka99} Omodaka, T., Maeda, T., Miyoshi, M., et al.\ 1999, \pasj, 51, 333 




\bibitem[Persi et al.(1998)]{Persi98} Persi, P., Tapia, M., Felli, M., Lagage, P.~O., \& Ferrari-Toniolo, M.\ 1998, \aap, 336, 1024 




\bibitem[{cf. also Puga et al.(2006)}]{Puga06} 
{Puga, E., Feldt, M., Alvarez, C., et al.\ 2006, \apj, 641, 373 }

\bibitem[Qiu et al.(2011)]{Qiu11} 
{ Qiu, K., Wyrowski, F., Menten, K.~M., et al.\ 2011, \apjl, 743, L25 }

\bibitem[Qiu \& Zhang(2009)]{Qiu09} 
{ Qiu, K., \& Zhang, Q.\ 2009, \apjl, 702, L66 }

\bibitem[Rabli \& Flower(2010)]{Rabli10} Rabli, D., \& Flower, D.~R.\ 2010, \mnras, 406, 95 



\bibitem[Reid et al.(2014)]{Reid14} Reid, M.~J., Menten, K.~M., Brunthaler, A., et al.\ 2014, \apj, 783, 130 

\bibitem[Reid \& Moran(1981)]{Reid81} Reid, M.~J., \& Moran, J.~M.\ 1981, \araa, 19, 231 

\bibitem[Ridge \& Moore(2001)]{Ridge01} 
{ Ridge, N.~A., \& Moore, T.~J.~T.\ 2001, \aap, 378, 495 }

\bibitem[Rodr{\'{\i}}guez-Kamenetzky et al.(2017)]{RodriguezKamenetzky} { Rodr{\'{\i}}guez-Kamenetzky, A., Carrasco-Gonz{\'a}lez, C., Araudo, A., et al.\ 2017, \apj, 851, 16 }

\bibitem[Rodr{\'{\i}}guez-Garza et al.(2017)]{Rodriguez17} Rodr{\'{\i}}guez-Garza, C.~B., Kurtz, S.~E., G{\'o}mez-Ruiz, A.~I., et al.\ 2017, \apjs, 233, 4


\bibitem[Rodriguez et al.(1982)]{Rodriguez82} Rodriguez, L.~F., Canto, J., \& Moran, J.~M.\ 1982, \apj, 255, 103 

\bibitem[Safron et al.(2015)]{Safron15} Safron, E.~J., Fischer, W.~J., Megeath, S.~T., et al.\ 2015, \apjl, 800, L5 




\bibitem[Sanna et al.(2017)]{Sanna17} Sanna, A., Moscadelli, L., Surcis, G., et al.\ 2017, \aap, 603, A94 

\bibitem[Sault \& Sowinski(2013)]{Sault13} Sault, R.J. \& Sowinski, K.\ 2013, EVLA Memo 169

\bibitem[Sch{\"o}ier et al.(2010)]{Schoier10} Sch{\"o}ier, F., van der Tak, F., van Dishoeck, E., \& Black, J.\ 2010, Astrophysics Source Code Library, ascl:1010.077 


\bibitem[Shirley(2015)]{Shirley15} Shirley, Y.~L.\ 2015, \pasp, 127, 299 

\bibitem[Sobolev et al.(1997)]{Sobolev97} Sobolev, A.~M., Cragg, D.~M., \& Godfrey, P.~D.\ 1997, \aap, 324, 211 


\bibitem[Stahler(2010)]{Stahler10} Stahler, S.~W.\ 2010, \mnras, 402, 1758 


\bibitem[Stutz et al.(2013)]{Stutz13} Stutz, A.~M., Tobin, J.~J., Stanke, T., et al.\ 2013, \apj, 767, 36 

\bibitem[Szymczak et al.(2017)]{Szymczak17} Szymczak, M. et al., 2017, MNRAS, submitted



\bibitem[Tapia et al.(2015)]{Tapia15} Tapia, M., Roth, M., \& Persi, P.\ 2015, \mnras, 446, 4088 

\bibitem[Tapia et al.(1996)]{Tapia96} 
{ Tapia, M., Persi, P., \& Roth, M.\ 1996, \aap, 316, 102 }


\bibitem[Thi et al.(2010)]{Thi10} 
{ Thi, W.-F., van Dishoeck, E.~F., Pontoppidan, K.~M., \& Dartois, E.\ 2010, \mnras, 406, 1409 }


\bibitem[Tolmachev(2011)]{Tolmachev11} Tolmachev, A.\ 2011, The Astronomer's Telegram, 3177


\bibitem[Urquhart et al.(2013)]{Urquhart13} Urquhart, J.~S., Moore, T.~J.~T., Schuller, F., et al.\ 2013, \mnras, 431, 1752 




\bibitem[Volvach et al.(2017b)]{Volvach17b} Volvach, A.~E., Volvach, L.~N., MacLeod, G., et al.\ 2017b, The Astronomer's Telegram, 10853  

\bibitem[Volvach et al.(2017a)]{Volvach17a} 
{ Volvach, A.~E., Volvach, L.~N., MacLeod, G., et al.\ 2017a, The Astronomer's Telegram, 10728}

\bibitem[Walsh et al.(1999)]{Walsh99} 
{ Walsh, A.~J., Burton, M.~G., Hyland, A.~R., \& Robinson, G.\ 1999, \mnras, 309, 905 }

\bibitem[Walsh et al.(1998)]{Walsh98} Walsh, A.~J., Burton, M.~G., Hyland, A.~R., \& Robinson, G.\ 1998, \mnras, 301, 640 



\bibitem[Willis et al.(2013)]{Willis13} 
{ Willis, S., Marengo, M., Allen, L., et al.\ 2013, \apj, 778, 96 }

\bibitem[Wu et al.(2014)]{Wu14} Wu, Y.~W., Sato, M., Reid, M.~J., et al.\ 2014, \aap, 566, A17 

\bibitem[Yen et al.(1969)]{Yen69} Yen, J.~L., Zuckerman, B., Palmer, P., \& Penfield, H.\ 1969, \apjl, 156, L27 

\bibitem[Yoo et al.(2017)]{Yoo17} Yoo, H., Lee, J.-E., Mairs, S., et al.\ 2017, \apj, 849, 69

\bibitem[Zernickel et al.(2012)]{Zernickel12} 
Zernickel, A., Schilke, P., Schmiedeke, A., et al.\ 2012, \aap, 546, A87 

\bibitem[Zhang et al.(2014)]{Zhang14} 
{ Zhang, Q., Qiu, K., Girart, J.~M., et al.\ 2014, \apj, 792, 116 }


\bibitem[Zuckerman et al.(1972)]{Zuckerman72} Zuckerman, B., Yen, J.~L., Gottlieb, C.~A., \& Palmer, P.\ 1972, \apj, 177, 59 

\end{thebibliography}
\end{document}